\documentclass[review]{elsarticle}

\usepackage{amsmath, amssymb}
\usepackage{lineno,hyperref}
\usepackage{siunitx}
\usepackage{graphicx}
\usepackage{caption}
\usepackage{subcaption}
\usepackage{algorithm}
\usepackage{algpseudocode}
\usepackage{algorithmicx}
\usepackage{booktabs, ltablex, makecell}
\newcolumntype{C}{>{\centering\arraybackslash}X}

\setcellgapes{3pt}

\algrenewcommand\algorithmicrequire{\textbf{Input:}}
\algrenewcommand\algorithmicensure{\textbf{Output:}}

\journal{Engineering Applications of Artificial Intelligence}

\begin{document}

\begin{frontmatter}

\title{Fault Detection in Induction Motors using Functional Dimensionality Reduction Methods}

\author[mainaddressSpain]{María Barroso}
\ead{maria.barrosoh@estudiante.uam.es}

\author[mainaddressArgentina]{José M. Bossio}
\ead{jbossio@ing.unrc.edu.ar}

\author[mainaddressSpain]{Carlos M. Alaíz}
\ead{carlos.alaiz@uam.es}

\author[mainaddressSpain]{\'{A}ngela Fern\'{a}ndez\corref{mycorrespondingauthor}}
\cortext[mycorrespondingauthor]{Corresponding author}
\ead{a.fernandez@uam.es}

\address[mainaddressSpain]{Departamento de Ingenier\'{i}a Inform\'{a}tica, Universidad Aut\'{o}noma de Madrid, Spain}
\address[mainaddressArgentina]{IITEMA-CONICET, Fac. de Ingeniería, Universidad Nacional de Río Cuarto, Argentina}

\begin{abstract}
The implementation of strategies for fault detection and diagnosis on rotating electrical machines is crucial for the reliability and safety of modern industrial systems.
The contribution of this work is a methodology that combines conventional strategy of Motor Current Signature Analysis with functional dimensionality reduction methods, namely Functional Principal Components Analysis and Functional Diffusion Maps, for detecting and classifying fault conditions in induction motors.
The results obtained from the proposed scheme are very encouraging, revealing a potential use in the future not only for real-time detection of the presence of a fault in an induction motor, but also in the identification of a greater number of types of faults present through an offline analysis.
\end{abstract}


\end{frontmatter}

\nolinenumbers

\section{Introduction}
\label{sec:intro}

The implementation of strategies for incipient faults detection and diagnosis on Rotating Electrical Machines (REM) is very important for the reliability and safety of modern industrial systems.
Its execution allows planning interruptions of continuous production processes in scheduled stops, thus reducing maintenance time and associated economic losses~\cite{mobley2004maintenance, etde_21044166}. 

The diagnosis of faults present in a REM is integrated by the detection, identification and isolation of an anomaly, which can be achieved by using the information obtained on the state of operation of the equipment or drive~\cite{Tavner1621470}.
As a result, it is possible to consider fault diagnosis as a pattern recognition problem with respect to the condition of a REM~\cite{1546063}.
To effectively diagnose faults in a REM, it is essential to distinguish between failures originating from the machine itself, whether electrical or mechanical, and those corresponding to the associated load~\cite{GANGSAR2020106908}.

In recent decades, with the advancement of communication technologies and the inclusion of control devices in REM, non-invasive faults detection and diagnosis techniques based on the use of electrical variables have been studied more than those that use acoustic emissions, analysis lubrication, thermography and vibrations.
The latter have been the techniques most widely used for some time, in which different methods are used for analysis, among the most common, Fast Fourier Transform (FFT) in the frequency domain, and wavelet analysis and empirical model decomposition in the domain time--frequency~\cite{LIU201833}.

The techniques based on electrical variables have focused mainly on those methods sustained on the Motor Current Signature Analysis (MCSA), instantaneous power analysis, and Park's vector analysis, among others~\cite{mobley2004maintenance}.
In this way it is possible to detect a large number of failures in induction motors, which are associated with the presence of certain components of the frequency spectrum.
It should be noted that, when performing these types of analysis, the presence of an expert in the area is required to carry out the task according to the information contained in the processed signals.

Its increasing use is due to the fact that the monitoring of the electrical variables to be analyzed is carried out without modifying the state or structure of the electrical machine~\cite{Tavner1621470}.
In particular, the sensors are placed on a control panel, thus avoiding problems with difficult access to the equipment to be analyzed and even reducing the risks for the operator in dangerous environments.

In an ideal scenario, all this process should be done in an automatic way.
Although machine learning methods fit nicely for this purpose, we should taken into account that it is very difficult to obtain a targeted sample in this context.
Thus, supervised models should be discarded.
As an alternative, dimensionality reduction and clustering methods could help to analyze and group the available electrical motor signals.

It is worth mentioning that electrical variables are functions, and hence they can be studied from a functional perspective using Functional Data Analysis (FDA;~~\cite{wang2016functional}) techniques.
The novelty of this article resides precisely on applying functional dimensionality reduction methods, namely Functional Principal Component Analysis (FPCA;~~\cite{Shang2014ASO}) and Functional Diffusion Maps (FDM;~\cite{barroso2023functional}), to detect and identify faults due to broken bars and low-frequency load oscillations in IMs.
Moreover, a complete methodology is proposed, covering from data processing to fault detection and fault diagnosis.

The rest of the paper is organized as follows.
In Section~\ref{sec:MLandIM} the state of the art in fault detection applying machine learning techniques is presented.
Section~\ref{sec:FDR} briefly reviews the theory under the functional dimensionality reduction techniques used in this work.
Section~\ref{sec:Exps} explains the data collection and the applied experimental methodology, and shows the obtained results.
Finally, Section~\ref{sec:Concl} presents some conclusions of this work, as well as possible research lines to further extend it.

\section{Machine learning and induction motor fault detection}
\label{sec:MLandIM}

In order to automate faults detection and diagnosis, significant progress has been made in the use of data processing, specifically data mining based on Artificial Intelligence methods.
This has been achieved through the implementation and integration of different machine learning techniques and computational statistics, as exposed in recent literature~\cite{GANGSAR2020106908, LIU201833, ReviewAlgorithmInductionMotors, app11062761, MethodsConditionDetectionMachines, BRITO2022108105, AdvancedSignalConditionMonitoring}.
These advancements enable non-experts in the field to analyze systems without detailed knowledge of the model being studied, resulting in simpler diagnoses~\cite{Tavner1621470}.

In~\cite{app11062761}, promising diagnostic techniques based on machine learning are presented, with a focus on their attributes.
In~\cite{MethodsConditionDetectionMachines}, an analysis of the advantages and disadvantages of various intelligent diagnostic techniques used in REM is presented, including decision trees, support vector machines, principal component analysis, and genetic algorithms.
This analysis was carry out based on the most common mechanical and electrical failures observed in these machines.

As explained in Section~\ref{sec:intro}, we will focus on unsupervised machine learning methods, specifically on dimensionality reduction techniques.
Regarding this field, in~\cite{AdvancedSignalConditionMonitoring} Principal Component Analysis (PCA) is identified as one of the most promising Machine Learning technique and is highlighted as a method that provides interesting results.
Its use allows the identification of the most significant failure characteristics and the extraction of underlying patterns, all while reducing data dimension.

Several case studies have focused on the use of PCA together with with various Machine Learning techniques for diagnosing REM failures, with particular emphasis on the diagnosis of broken bars.
Such is the case of~\cite{DeterminationBrokenPCA}, where an advanced signal processing method based on wavelet analysis, PCA, and multi-layer neural networks is presented.
This technique enables the extraction of suitable characteristics, reduces the correlation between the extracted features, and determines the magnitude of a failure due to broken bars in an IM.

On the other hand, \cite{7544203} provides a comparative analysis of three methods for detecting broken bars in an induction motor, based on the electrical signal analysis, particularly MCSA, MSCSA and PCA.
Additionally, \cite{7049012} presents a method for the detection of broken bars through the use of PCA in the three stator currents, to later be used in the calculation of $\mathcal{Q}$ statistic that determines the presence or absence of the fault.

As mentioned before, due to its nature, electrical variables can be studied from a functional perspective using FDA techniques that, as far as we have read, no study has used before for trying to solve the problem at hand.

\section{Functional dimensionality reduction methods}
\label{sec:FDR}

Dimensionality reduction methods are statistical techniques where high-dimensional data are represented in lower dimensional spaces, for example by capturing most of the variance of the original data in the new space, like PCA does, or by reflecting lower dimensional underlying manifolds where the original data lay, as manifold learning techniques intend~\cite{manifold_learning}.

When data are functions, they live in infinite-dimensional spaces, and hence finding low-dimensional representations becomes essential.
Finding reliable representations of low-dimensional data, specifically in two or three dimensions, is beneficial in real-world problems for visualization, description and general exploration purposes~\cite{Unsupervised_Functional_Data_Analysis}.
Moreover, these representations can be used as feature vectors in supervised machine learning algorithms which require multivariate inputs~\cite{Benchmarking_time_series_classification}.

The most popular technique is FPCA. However, non-linear dimensionality reduction methods such as FDM and Isomap~\cite{Nonlinear_manifold_representations_for_functional_data} have been gaining popularity in recent years and have outperformed FPCA in some data applications.

In the next subsections, we will briefly introduce FDA, and present the theoretical framework for FPCA and FDM.

\subsection{Functional data analysis}

Functional Data Analysis ~\cite{wang2016functional} studies samples of functions $x_1(t), \dotsc, x_N(t)$, where $t\in \mathcal{J}$, $\mathcal{J}$ is a compact interval, and each $x_i(t)$ is the observation of an independent functional variable $X_i$ identically distributed as $X$. 
It is usual to assume that the functional variable $X$ is a second order stochastic process, $\operatorname{E}[X^2]<\infty$, and takes values in the Hilbert space of square integrable functions $L^2([a,b])$ defined on the closed interval $[a,b] \subset \mathbb{R}$.
Square integrable functions form a vector space and we can define the inner product of two functions by $\langle x,y \rangle = \int_{a}^b x(t) y(t) dt$.
The inner product allows to introduce the notion of distance between functions by the $L^2$-norm $\|x\|^2 = \langle  x,x \rangle = \int_{a}^b x^2(t)dt.$
Therefore, in $L^2$ space, the distance between two functions can be calculated as the norm of their difference, which is expressed as $\|x-y\|$.

\subsection{Functional PCA}
\label{subsec:FPCA}

Functional Principal Component Analysis ~\cite{Shang2014ASO}  is a linear functional dimensionality reduction method that generalizes multivariate Principal Component Analysis \cite{abdi2010principal} to the case where data are functions. 
In FPCA, the infinite-dimensional random functions are projected onto the lower dimensional subspace generated by the eigenfunctions of the covariance operator.

Let $x_1(t), \dotsc, x_N(t)$ be the realizations of a stochastic processes over a compact domain.
The sample variability is characterized by the spectral decomposition of the sample covariance operator, $\hat{\Gamma} y(t) = \int_a^b \hat{\gamma}(t,s)y(s)ds$ where $\hat{\gamma}(s,t)= N^{-1} \sum_{i=1}^N x_i(s)x_i(t)$ is the sample covariance function.
The directions $\xi_l$ of the FPCA projection into an $L$-dimensional subspace are chosen such that they maximized the variance of the projection; more specifically they are the solution of the following problem:

\begin{equation*}
    \max_{\xi_l} \widehat{\mathrm{Var}}[\langle \xi_l, X\rangle] \;\mbox{s.t.}\; \langle \xi_l, \xi_k \rangle = \delta_{lk},\; k\leq l,\; l=1,...,L.
\end{equation*}

The above expression can be simplified by using the sample covariance operator defined as
\begin{equation*}
    \max_{\xi_l} \langle \hat{\Gamma} \xi_l, \xi_l \rangle \;\mbox{s.t.}\; \langle \xi_l, \xi_k \rangle = \delta_{lk},\; k\leq l,\; l=1,...,L.
\end{equation*}
The solutions of this problem are obtained by solving the eigenequation 
\begin{equation}\label{eq:spectral_fpca}
    \hat{\Gamma} \xi_l (t) = \lambda_l \xi_l (t), \; t\in[a,b],
\end{equation}
where $\lambda_1 \geq \lambda_2 \geq\cdots \geq 0$ are the eigenvalues and $\xi_1,\xi_2\dots$ are the eigenfunctions, which form an orthonormal basis for the space of functions being analyzed.
Hence, $\hat{x}_i = \sum_{l=1}^L \hat{\theta}_{li}\xi_l$, with $\hat{\theta}_{li}= \langle \xi_l, x_i\rangle$ is a good approximation of $x_i$ for a relevant choice of $L$.

To apply this method, there are two possible strategies to approach the eigenanalysis problem~\eqref{eq:spectral_fpca}: discretizing the functions or expressing the functions in a known basis.
In both cases, we convert the continuous functional eigenanalysis problem into an approximately equivalent matrix eigenanalysis task. 
The whole procedure of the first strategy, which will be the one used in the experiments, is shown in Algorithm~\ref{alg:fpca_discretized}. 
Here, FPCA is equivalent to a standard multivariate PCA with the metric defined by the quadrature weight matrix. 
For more information about the other strategy, please refer to~\cite{Ramsay_FDA}.

\begin{algorithm}
\caption{FPCA over discretized functions.}\label{alg:fpca_discretized}
\begin{algorithmic}[1]
\Require
\Statex $\mathrm{X}$ -- Functional data matrix
\Statex $L$ -- Embedding dimension
\Statex $\{t_j\}_{j=1}^M$ -- Quadrature points
\Ensure
\Statex $\{\tilde{\xi}_l\}_{l=1}^L$ -- Discretized principal component functions
\Statex $\{\hat{\theta}_{l}\}_{l=1}^{L}$ -- Principal component scores \\
Compute sample covariance matrix $\hat{\Sigma} = N^{-1}\mathrm{X}^{\top}\mathrm{X}$, where $N$ is the number of functional data.\\
Compute weight matrix $\mathrm{W}$ from quadrature weights using some numeric integration rule. \\
Obtain eigenvalues $\{\lambda_l\}_{l=1}^L$ and eigenvectors $\{\mathrm{u}_l\}_{l=1}^L$ of $\mathrm{W}^{1/2}\hat{\Sigma}\mathrm{W}^{1/2}$ satisfying
\begin{equation*}
\mathrm{W}^{1/2}\hat{\Sigma}\mathrm{W}^{1/2} \mathrm{u}_l = \lambda_l \mathrm{u}_l \; \mbox{s.t.}\; \mathrm{u}_l^{\top}\mathrm{u}_k = \delta_{lk}, \; l,k=1,\dotsc,L.
\end{equation*}
\\
Calculate discretized principal component functions $\tilde{\xi}_l = \mathrm{W}^{-1/2}\mathrm{u}_l$.\\
Calculate principal components scores
$\hat{\theta}_{l} = \mathrm{X}\mathrm{W}\tilde{\xi}_l$.
\end{algorithmic}
\end{algorithm}
\bigskip

\subsection{Functional DM}
\label{subsec:FDM}

Functional Diffusion Maps \cite{barroso2023functional} is a nonlinear dimensionality reduction algorithm applied to functional data that extend multivariate Diffusion Maps (DM,~ \cite{Diffusion_Maps, An_Introduction_to_Diffusion_Maps}) to the functional domain.
FDM seeks to identify low-dimensional representations of $L^2$ functions on nonlinear functional manifolds after defining a diffusion process on a normalized graph based on pairwise similarities between functional data.

In more detail, let $\mathcal{X} = {x_1(t), ..., x_N(t)}$ be the realizations of a stochastic process over a compact domain.
In this context, $\mathcal{X}$ is assumed to lie on a functional manifold $\mathcal{M} \subset L^2([a,b])$.
To identify the underlying manifold, the initial stage of FDM involves building a weighted graph $\mathrm{G}=(\mathcal{X}, \mathrm{K})$, where the graph vertices are functions $x_i(t)$ and the weights $k_{ij}$ are obtained from a symmetric and pointwise positive $N\times N$ kernel matrix $\mathrm{K}$.
This kernel matrix defines a local measure of similarity within a certain neighborhood, so that outside the neighborhood the function quickly goes to zero.
The standard kernel used to compute the similarity between functional data is the Gaussian kernel, $k_{ij} = \exp{\frac{-\|x_i(t)-x_j(t)\|^2_{ L^2}}{2\sigma^2}}$, where the size of the local neighborhood considered is determined by the hyperparameter $\sigma$. Alternatively, the Laplacian kernel can be used, defined as $ k_{ij} = \exp{\frac{-\|x_i(t)-x_j(t)\|_{L^1}}{\sigma^2}}$. These types of kernels define a connected graph.

Once the matrix $\mathrm{K}$ is obtained, the connected graph is normalized using a density parameter $\alpha\in [0,1]$.
This results in a new graph, denoted as $\mathrm{G}'=(\mathcal{X},\mathrm{K}^{(\alpha)})$, where the entries of $\mathrm{K}^{(\alpha)}$ are $k_{ij}^{(\alpha)}=\frac{k_{ij}}{d_i^\alpha d_j^\alpha}$.
Here, $d_i=\sum_{j=1}^N k_{ij}$ is the degree of the graph and the power $d_i^\alpha$ approximates the density of each vertex.  

Now, we can create a Markov chain on the normalized graph whose transition matrix $\mathrm{P}$ is defined by $p_{ij} = \frac{k_{ij}^{(\alpha)}}{d_i^{(\alpha)}}$, where $d_i^{(\alpha)}=\sum_j k_{ij}^{(\alpha)}$.
The transition matrix $\mathrm{P}$ provides the probabilities of arriving from node $i$ to node $j$ in a single step.
By taking powers of the $\mathrm{P}$ matrix, we can increase the number of steps taken in the random walk.
This defines a \textit{diffusion process} that reveals the global geometric structure of $\mathcal{X}$ at different scales.

Now we are ready to define a diffusion distance $\mathrm{D_{T}}$ based on the geometrical structure of the obtained diffusion process, 
\begin{equation*}
    \mathrm{D}_T^2 (x_i(t),x_j(t)) = \|p_{i\cdot}^T - p_{j\cdot}^T\|^2_{\mathrm{L}^2(\frac{1}{\pi})} = \sum_k \frac{\left(p_{ik}^T - p_{jk}^T\right)^2}{\pi_k}.
\end{equation*}
This metric measures the similarities between data as the connectivity or probability of transition between them.
Therefore, $T$ plays the role of a scale parameter and $\mathrm{D}_T^2 (x_i(t),x_j(t))$ will be small if there exist a lot of paths of length $T$  that connect $x_i(t)$ and $x_j(t)$.

Spectral analysis of the Markov chain allows us to consider an alternative formulation of the diffusion distance that uses eigenvalues and eigenvectors of $\mathrm{P}$.
As detailed in~\cite{Diffusion_Maps}, even though $\mathrm{P}$ is not symmetric, it makes sense to perform its spectral decomposition using its left and right eigenvectors, such that $p_{ij}=\sum_{l\geq 0} \lambda_l (\psi_l)_i (\varphi_l)_j$.

The eigenvalue $\lambda_0=1$ of $\mathrm{P}$ is discarded  since $\psi_0$ is a vector with all its values equal to one.
The other eigenvalues $\lambda_1, \lambda_2, \dotsc,\lambda_{N-1}$ tend to $0$ and satisfy $\lambda_l<1$ for all $l \geq 1$.
Thus, the diffusion distance can be approximated by the first $L$ eigenvalues and eigenvectors using the new representation of $\mathrm{P}$, $\mathrm{D}_T^2(x_i(t), x_j(t)) \approx \sum_{l=1}^{L}\lambda_l^{2T} \left((\psi_l)_i - (\psi_l)_j\right)^2.$

Finally, the diffusion map is given by
\begin{align}
    \Psi_T(x_i(t)) = \begin{pmatrix}
           \lambda_1^T \psi_1(x_i(t)) \\
           \vdots \\
           \lambda_{L}^T \psi_{L}(x_i(t))
         \end{pmatrix},
\end{align}
satisfying that the diffusion distance on the original space can be approximated by the Euclidean distance of the $\Psi_T$ projections in $\mathbb{R}^{L}$:
\begin{equation*}
    \mathrm{D}_T^2(x_i(t), x_j(t)) \approx \sum_{l=1}^{L}\lambda_l^{2T} \left((\psi_l)_i - (\psi_l)_j\right)^2 = \|\Psi(x_i(t)) - \Psi(x_j(t))\|^2.
\end{equation*}

The complete methodology of the method is presented in Algorithm~\ref{alg:fdm}.

\begin{algorithm}
\caption{FDM.}\label{alg:fdm}
\begin{algorithmic}[1]
\Require
\Statex $L$ -- Embedding dimension
\Statex $T$ -- Steps in random walk
\Statex $\alpha$ -- Density parameter
\Statex $\mathcal{K}$ -- Kernel operator $\mathcal{K}:L^2([a,b])\times L^2([a,b])\to \mathbb{R}$
\Statex $\mathcal{X}  = \{x_1(t), \dots, x_N(t)\}$ -- Functional dataset
\Ensure
\Statex $\Psi_T(\mathcal{X} )$ -- Embedded functional data \\
Construct $\mathrm{G}=(\mathcal{X} , \mathrm{K})$, where $\mathrm{K}$ is a positive and symmetric matrix with $k_{ij}= \mathcal{K}(x_i(t), x_j(t))$. \\
Compute density of each vertex: $d_i^{\alpha} = \left(\sum_{j=1}^N k_{ij}\right)^\alpha.$  \\
Construct $\mathrm{G}'=(\mathcal{X} , \mathrm{K}^{(\alpha)})$ normalized by $\alpha$ with $k_{ij}^{(\alpha)} = \frac{k_{ij}}{d_i^\alpha d_j^\alpha}.$  \\
Define transition matrix $\mathrm{P}$ with $p_{ij} = \frac{k_{ij}^{(\alpha)}}{d_i^{(\alpha)}}$, where $d_i^{(\alpha)}=\sum_j k_{ij}^{(\alpha)}.$  \\
Obtain eigenvalues $\{\lambda_l\}_{l\geq0}$ and right eigenvectors $\{\psi_l\}_{l\geq 0}$ of $\mathrm{P}$ such that 
        \[
          \begin{cases}
            1      &= \lambda_0 > \lambda_1 \geq \dotsc\\
            \mathrm{P}\psi_l &= \lambda_l \psi_l.
          \end{cases}
        \]
        \\
Calculate diffusion maps  $$\Psi_T(x_i(t)) = \begin{pmatrix}
           \lambda_1^T \psi_1(x_i(t)) \\
           \vdots \\
           \lambda_{L}^T \psi_{L}(x_i(t))
         \end{pmatrix} , \; \forall x_i(t)\in \mathcal{X}.$$
\end{algorithmic}
\end{algorithm}

\section{Experiments}
\label{sec:Exps}

\subsection{Data collection}
\label{subsec:data}

For this particular research, several tests were carried out in an experimental test bench made up of two induction motors mechanically coupled through a Gummi type flexible coupling.
There it was possible to reproduce the failure conditions due to broken bars and low-frequency load oscillation.

The motor under test is connected directly to the grid while the motor used as load is connected to a variable speed drive with torque control.
An oscillating torque reference of adjustable amplitude and frequency was added to the reference of said torque control, thus being able to represent the failure due to a low-frequency oscillating load. Specifically, at a frequency of \SI{1}{\hertz} and \SI{2}{\hertz}, with a percentage of the nominal load torque of the order of \SI{2}{\percent} and \SI{3}{\percent}, depending on each case.
In addition, the torque control allowed to adapt its operation according to the different load values in the analyzed IM, particularly with values of \num{0}, \num{20}, \num{40}, \num{60} and \SI{80}{\percent} load.

The condition of a healthy motor and those with interrupted or broken bars were represented in the motor under test due to the possibility of using different types of rotors, in particular a healthy rotor and three rotors with broken bars (\num{1}, \num{2} and \num{3} continuous bars).
In the tests carried out, two phase currents and two line voltages were measured, using AC current probes, and isolated voltage probes with $10:1$ attenuation, respectively.
Each measurement has a length of \num{128000} samples and a sampling frequency of \num{8000} samples/second.
More technical data and parameters of the IM for laboratory experimental results are shown in Table~\ref{TAB:technical-data}.

\begin{tabularx}{\linewidth}{@{}*{8}{c} }
    \toprule
\thead{Rated\\ power}
    & \thead{Rated\\ voltage}
        & \thead{Frequency}
            & \thead{Rated\\ current}
                & \thead{Rated\\ speed}
                    & \thead{Power\\ factor}
                        & \thead{Rotor\\ inertia}
                            & \thead{N. rotor\\ bars} \\
    \midrule
\endfirsthead
\SI{5.5}{\kilo\watt} & \SI{380}{\volt}  &  \SI{50}{\hertz} & \SI{11.1}{\ampere} & \SI{1470}{rpm} & \num{0.85} & \SI{0.02}{\kilogram\metre^2} & \num{40}\\
    \bottomrule
\caption{Technical data and parameters of the IM for laboratory experimental results.}
\label{TAB:technical-data}
\end{tabularx}

\begin{figure}[h]
    \centering
    \includegraphics[scale=0.35]{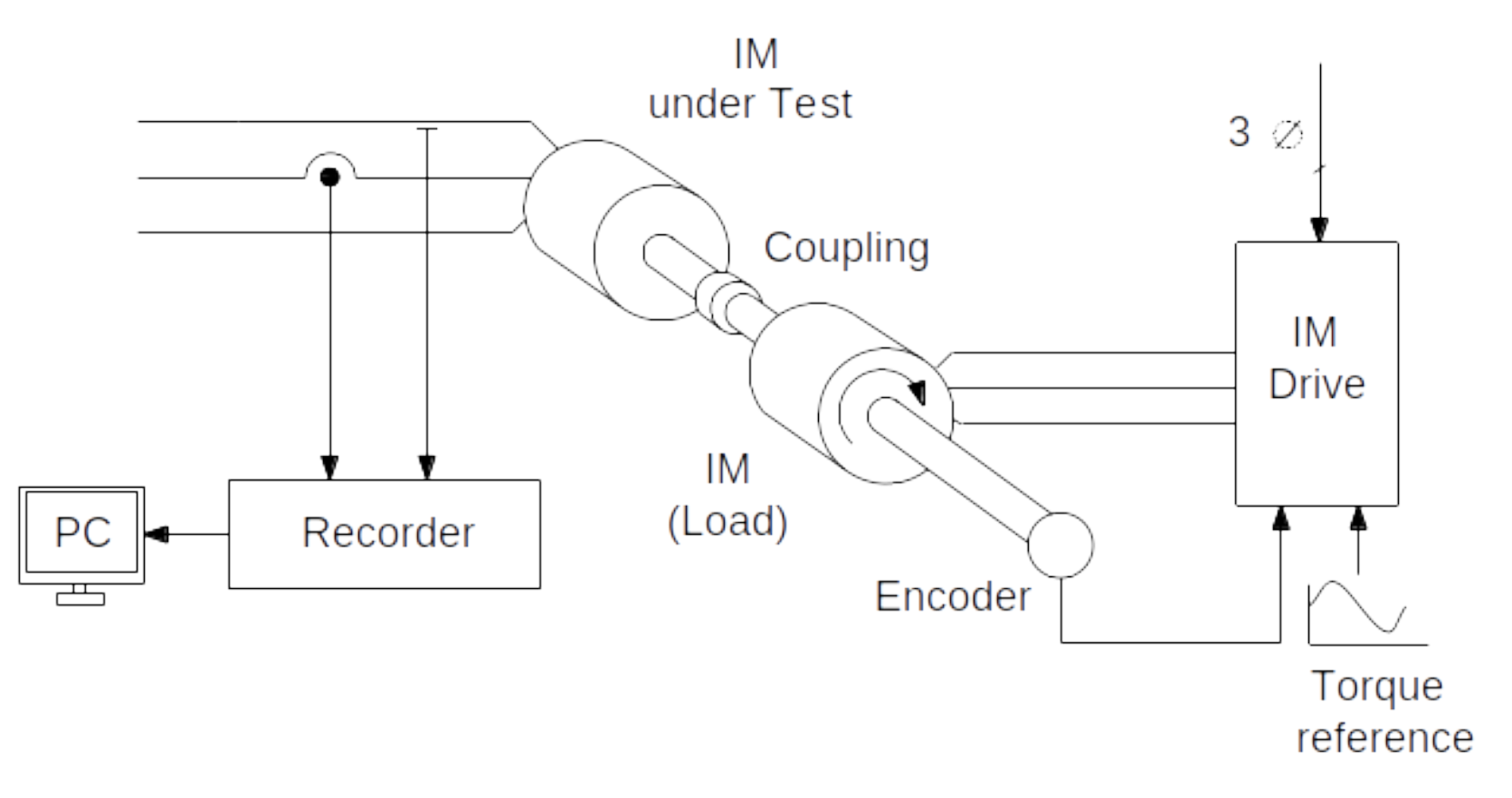}
    \caption{Schematic diagram of the test bench.}
    \label{FIG:diagram-bench}
\end{figure}

The recording and storage of electrical variables were carried out using a digital recorder with \num{4} channels, with data storage capacity through an internal memory in the recorder itself.
In Figure~\ref{FIG:diagram-bench}, a schematic diagram of the laboratory assembly is shown (test bench and measurement equipment).

The analyzed dataset is made up of \num{10} measurements done with the IM without anomalies (Healthy Motor, HM) per load percentage.
Each measurement was repeated twice per experimental setup, so a total of \num{50} HM measurements were compiled.
We have to add \num{10} measurements for the motor with \num{1} broken bar (1BB), \num{10} measurements with \num{2} broken bars (2BB), and \num{10} measurements with \num{3} broken bars (3BB), repeating again two measurements for each percentage of load in the three cases; they were also collected \num{10} measurements with load oscillation at \SI{1}{\hertz} with an amplitude of \SI{2}{\percent} of the nominal load torque (Sinusoidal Oscillation Signal, SS\_1), and \num{10} with \SI{3}{\percent} of the load torque, with two measurements for each load percentage in both cases; and finally, \num{20} measurements with load oscillation at \SI{2}{\hertz} with \SI{2}{\percent} and \SI{3}{\percentº} of the nominal load torque respectively (SS\_2), two per load percentage in each case, were obtained. 
In this way, a total of  \num{120} current measurements are available for data analysis using the proposed functional dimensionality reduction techniques. 

Apart from these current signals, together with the two line voltage signals mentioned above, instantaneous active power (IAP) signals are obtained for the cases of motors with faults, as in~\cite{Bossio5071292}.
Therefore, \num{70} IAP measurements are also available. 

To summarize, Table~\ref{TAB:faults-summary} shows the labels used for each type of induction motor diagnostic throughout our experiments and their frequencies.

\begin{table}[]
    \centering
    \small
    \begin{tabular}[t]{llcc}
    \toprule
    IM Condition & Tag & Current Freq & IAP Freq  \\
    \toprule
    Healthy Motor & HM & \num{50} & -\\
    \midrule
    Induction Motor with 1 Broken Bar & 1BB & \num{10} & \num{10}\\
      Induction Motor with 2 Broken Bars & 2BB & \num{10} & \num{10}\\
     Induction Motor with 3 Broken Bars & 3BB & \num{10} & \num{10}\\
     \midrule
    Sinusoidal Signal load at \SI{1}{\hertz}, \SI{1}{\milli\volt} & SS\_1\_A & \num{10} & \num{10} \\
      Sinusoidal Signal load at \SI{1}{\hertz}, \SI{1.5}{\milli\volt} & SS\_1\_B & \num{10} & \num{10}\\
     Sinusoidal Signal load at \SI{2}{\hertz}, \SI{1}{\milli\volt} & SS\_2\_A & 10 & 10\\
     Sinusoidal Signal load at \SI{2}{\hertz}, \SI{1.5}{\milli\volt} & SS\_2\_B  & \num{10} & \num{10}\\
    \bottomrule
    \end{tabular}
    \caption{Current and instantaneous active power data information.}
    \label{TAB:faults-summary}
\end{table}

\subsection{Experimental methodology}
\label{subsec:meth}

This section presents the experimental framework developed in this research, which uses functional dimensionality reduction methods to detect and diagnose faults in induction motors. 

The current and instantaneous active power data will be analyzed, including raw signals, their derivatives, and their Fourier transform. 
The latter are referred to as signatures in the literature and so here.
Functional Principal Component Analysis and Functional Diffusion Maps will be applied to each type of data, and the results obtained by each technique will be compared.
Finally, a scheme for detecting and diagnosing failures in IMs will be proposed based on the obtained results.

\subsubsection{Data preprocessing}
 
First of all, we align data using the first $x$-axis cut-off point.
Then, the datum corresponding to one broken bar with load at \SI{20}{\percent} was dropped as it was identified as an outlier by expert knowledge.
Therefore, this implies working with a total of \num{119} current signals and \num{69} IAP signals.
We scale data to the range of \num{-1} and \num{1}.
In this way, by normalizing the signals, it is possible to compare them independently of the associated load percentage.
Since the managed data are periodical, we just consider the first \num{750} steps as a representative sample.
Thus, we deal with the curves in the first \SI{9.3625}{\milli\second}.

Next, we estimate the derivatives of both current and IAP signals by finite differences~\cite{derivative_finite_difference} and the current and active power signatures by applying the Fast Fourier Transform (FFT;~\cite{fft}) method to the normalized data.

\begin{figure}[htp]
\centering
     \begin{subfigure}{0.48\textwidth}
         \centering
         \includegraphics[width=\textwidth,height=1.5in]{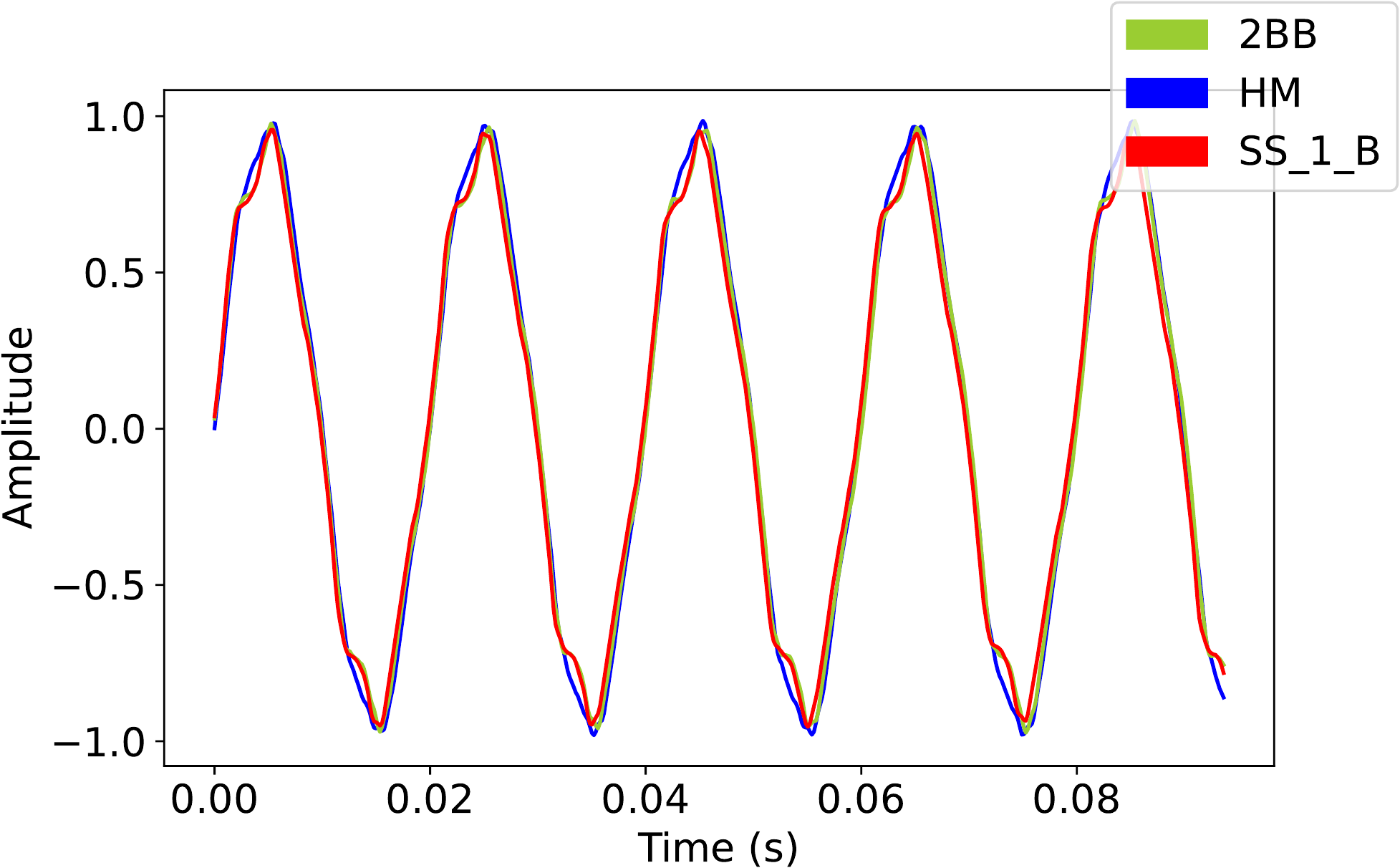}
         \caption{Current signals.}
     \end{subfigure}
     \hfill
     \begin{subfigure}{0.48\textwidth}
         \centering
         \includegraphics[width=\textwidth,height=1.5in]{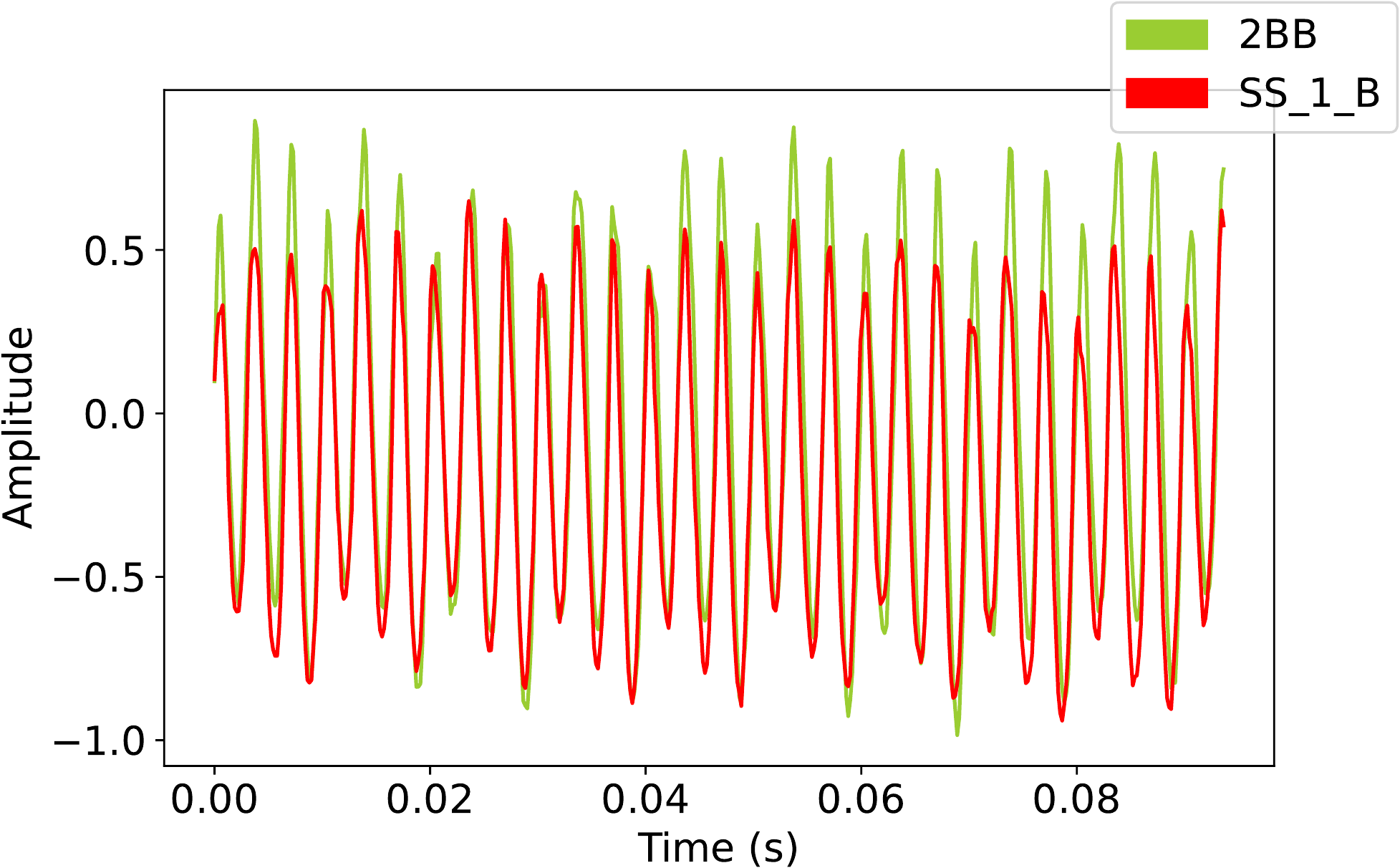}
         \caption{IAP signals.}
     \end{subfigure}
     \\
     \begin{subfigure}{0.48\textwidth}
         \centering
         \includegraphics[width=\textwidth,height=1.5in]{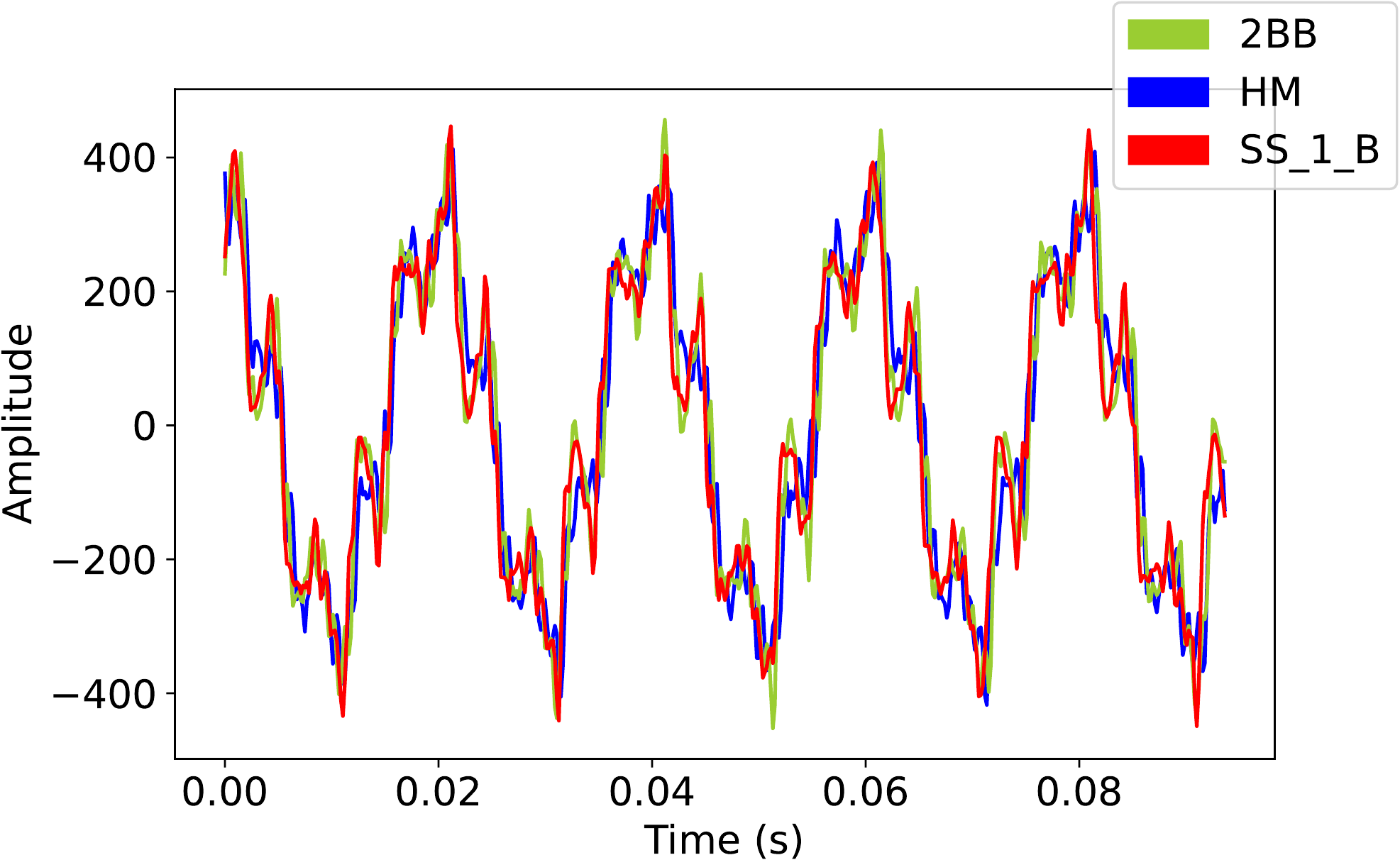}
         \caption{Current signal derivatives.}
     \end{subfigure}
     \hfill
     \begin{subfigure}{0.48\textwidth}
         \centering
         \includegraphics[width=\textwidth,height=1.5in]{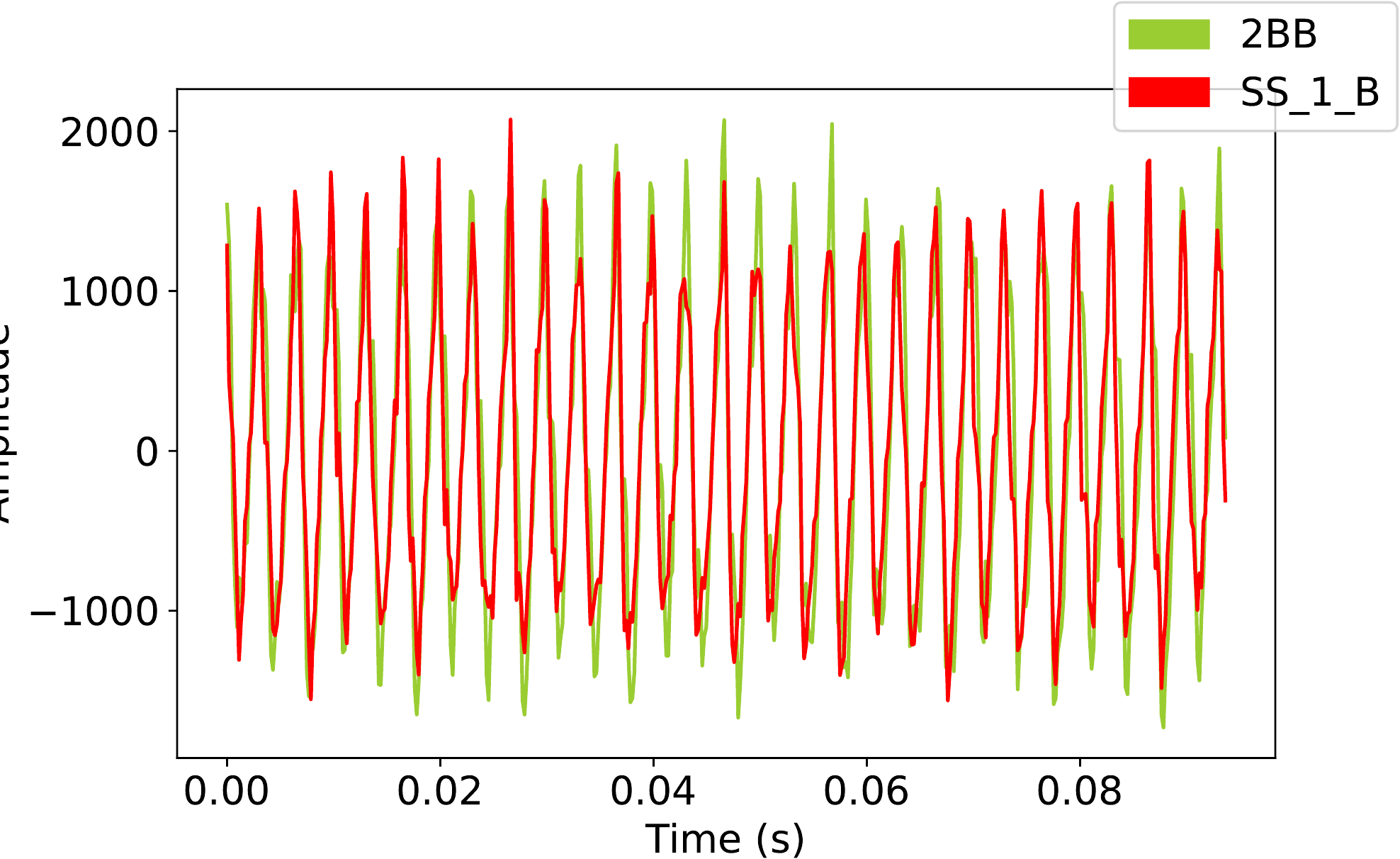}
         \caption{IAP signal derivatives.}
     \end{subfigure}
     \\
     \begin{subfigure}[b]{0.48\textwidth}
         \centering
         \includegraphics[width=\textwidth,height=1.5in]{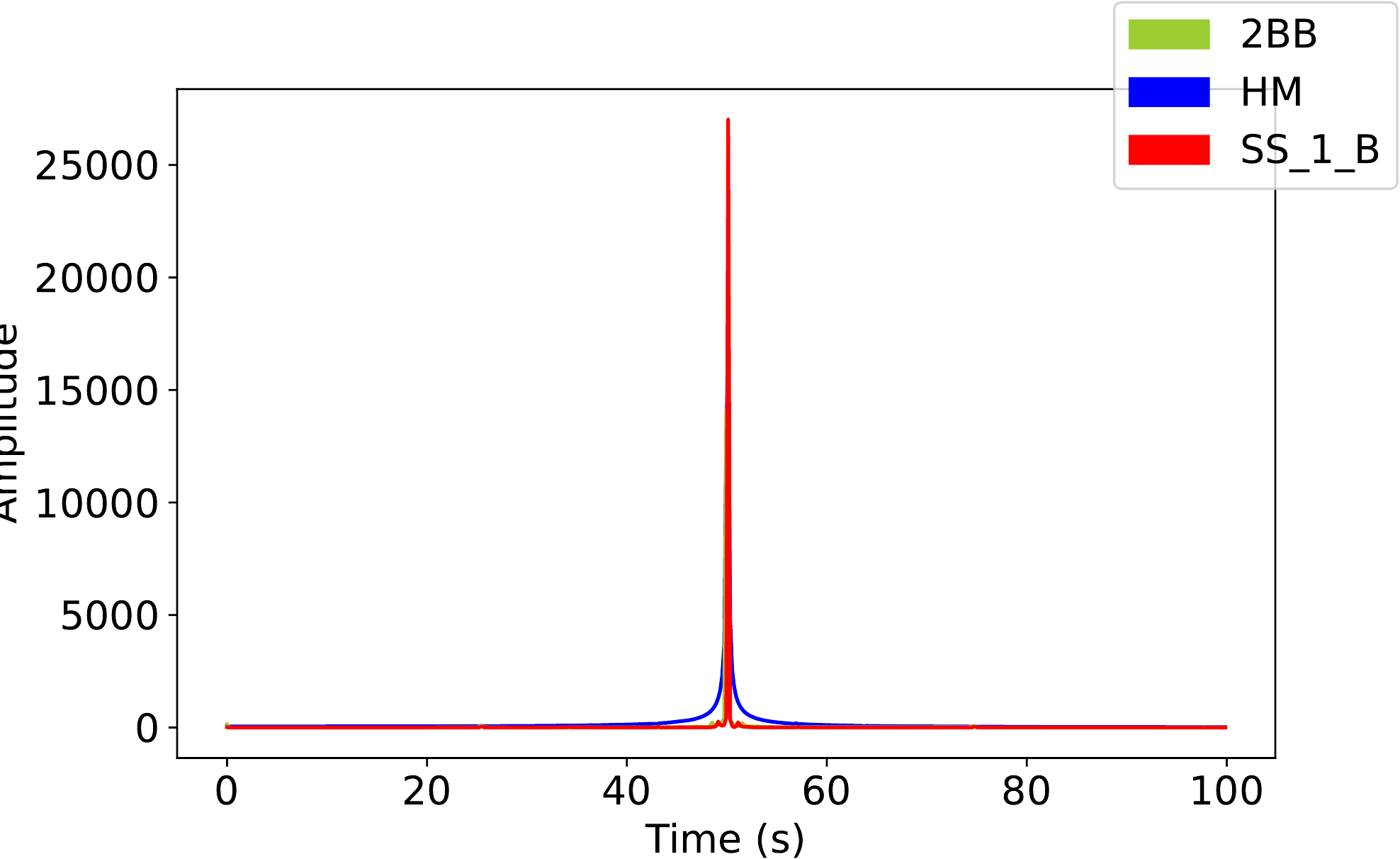}
         \caption{Current signatures.}
     \end{subfigure}
     \hfill
     \begin{subfigure}[b]{0.48\textwidth}
         \centering
         \includegraphics[width=\textwidth,height=1.5in]{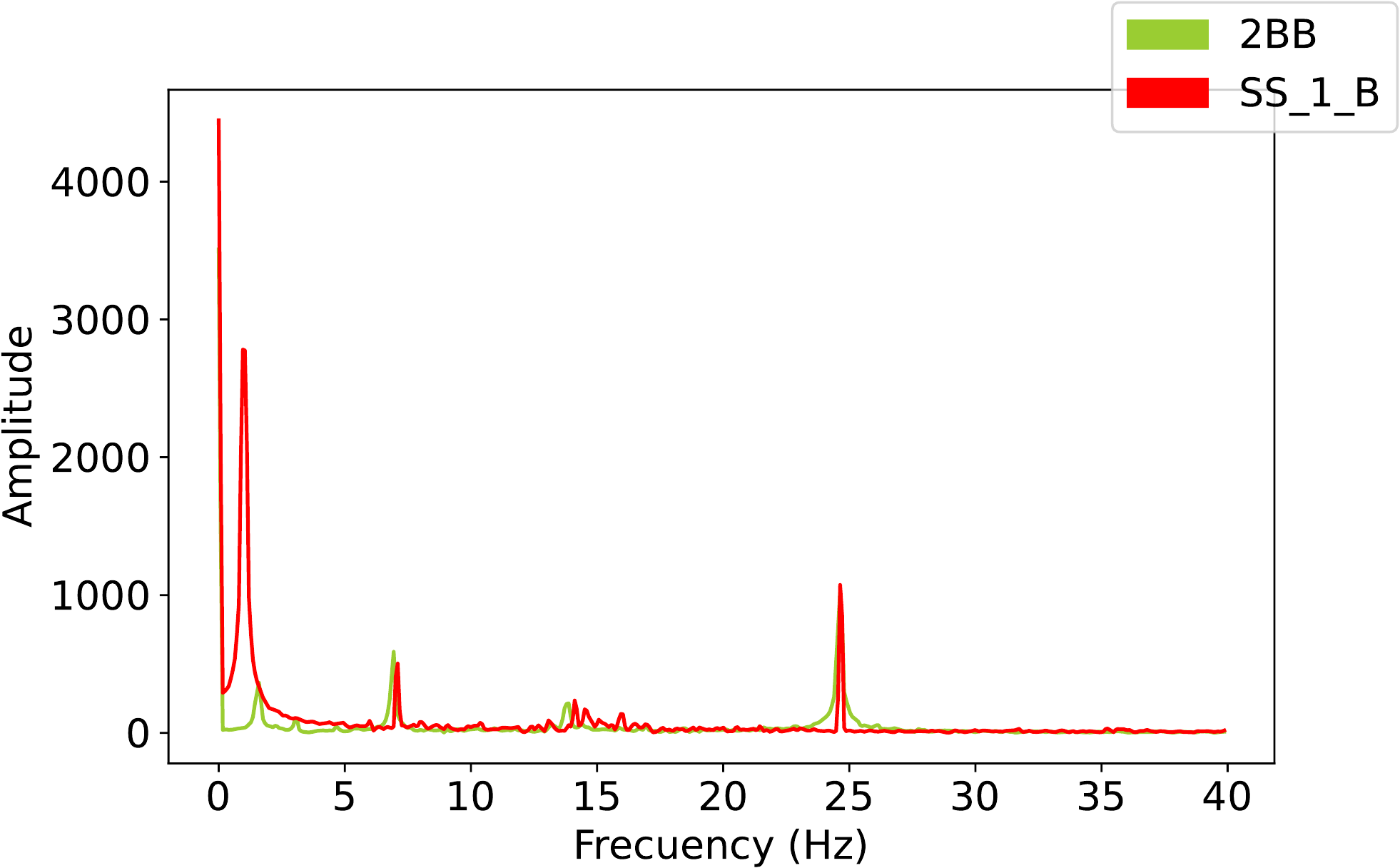}
         \caption{IAP signatures.}
     \end{subfigure}
    \caption{Current and IAP data from motors with faults due to two broken bars and motors with faults due to \SI{1}{\hertz}, \SI{1.5}{\milli\volt} load oscillations, using load at \SI{80}{\percent}.
    Current data from the healthy motor are also shown in the left panel.}
    \label{FIG:current-power-1bb}
\end{figure}

In Figure~\ref{FIG:current-power-1bb}, examples of current data and instantaneous active power data obtained from motors with faults due to broken bars and motors with low-frequency load oscillations are shown.
Specifically, these data were obtained from the motor with two broken bars and the motor with a \SI{1}{\hertz}, \SI{1.5}{\milli\volt} load oscillations, using load at \SI{80}{\percent}.
The figure shows both preprocessed signals, as well as derivatives of the signals and FFTs.
In addition, for the current data, signals obtained from the healthy motor at the same load percentage are also shown.

By visually analyzing the current data, we observe differences between data from faulty motors and data from the healthy motor.
These differences are clearer in the case of FFTs, which show peaks around the fundamental frequency, equal to \SI{50}{\hertz}, when motors have faults.

Analyzing the IAP data, we barely find differences in the preprocessed signals and derivatives that allow us to distinguish the type of motor fault.
However, considerable magnitude peaks are observed at low frequencies in the FFTs of the sinusoidal signals, which do not appear in the FFTs of motors with broken bars.

It is worth noting that when visually discriminating healthy motors from faulty motors using FFTs and discriminating motors with low-frequency load oscillations faults using FFTs of IAP, it is not necessary to resort to analyzing instantaneous reactive power (IRP).

\subsubsection{Experimental configuration}

FDM require an initial analysis to identify the suitable parameters to reveal patterns of interest or clusters in the data.
Table~\ref{TAB:fdm-hyperparameters} shows the hyperparameters obtained after a visual analysis for different parameters configurations. 
Based on the results, we can conclude that the Gaussian kernel gives better outcomes for current data, while the Laplacian kernel appears to be more appropriate for signatures.

\begin{table}[h]
    \centering
    \small
    \begin{tabular}[t]{llccc}
    \toprule
    Data & Type & Kernel & $\sigma$ & $\alpha$\\
    \toprule
    Current & Signal & Gaussian & \num{0.035} & \num{1.0} \\
    & Derivative & Gaussian & \num{10.0} & \num{0.0} \\
    & Signature & Laplacian & \num{100.0} & \num{1.0} \\
    \midrule
    IAP & Signal & Gaussian & \num{0.1} & \num{0.5}  \\
    & Derivative & Gaussian & \num{5.0} & \num{0.0} \\
    & Signature & Laplacian & \num{38.0} & \num{0.25}\\
    \bottomrule
    \end{tabular}
    \caption{FDM hyperparameters.}
    \label{TAB:fdm-hyperparameters}
\end{table}

\subsection{Experimental results}
\label{subsec:expresult}

In these experiments, the embeddings obtained from FPCA and FDM are analyzed for both current and instantaneous active power data in the time and frequency domains.
Initially, we will apply the functional dimensionality reduction techniques discussed previously to the raw current signal and its derivative.
Following this, we will apply these methods to the current signatures, and the same analysis will be repeated for the instantaneous active power.

\subsubsection{Analysis of the current signal and its derivative}

\begin{figure}[t]
\centering
     \begin{subfigure}[htp]{0.48\textwidth}
         \centering
         \includegraphics[width=\textwidth,height=1.5in]{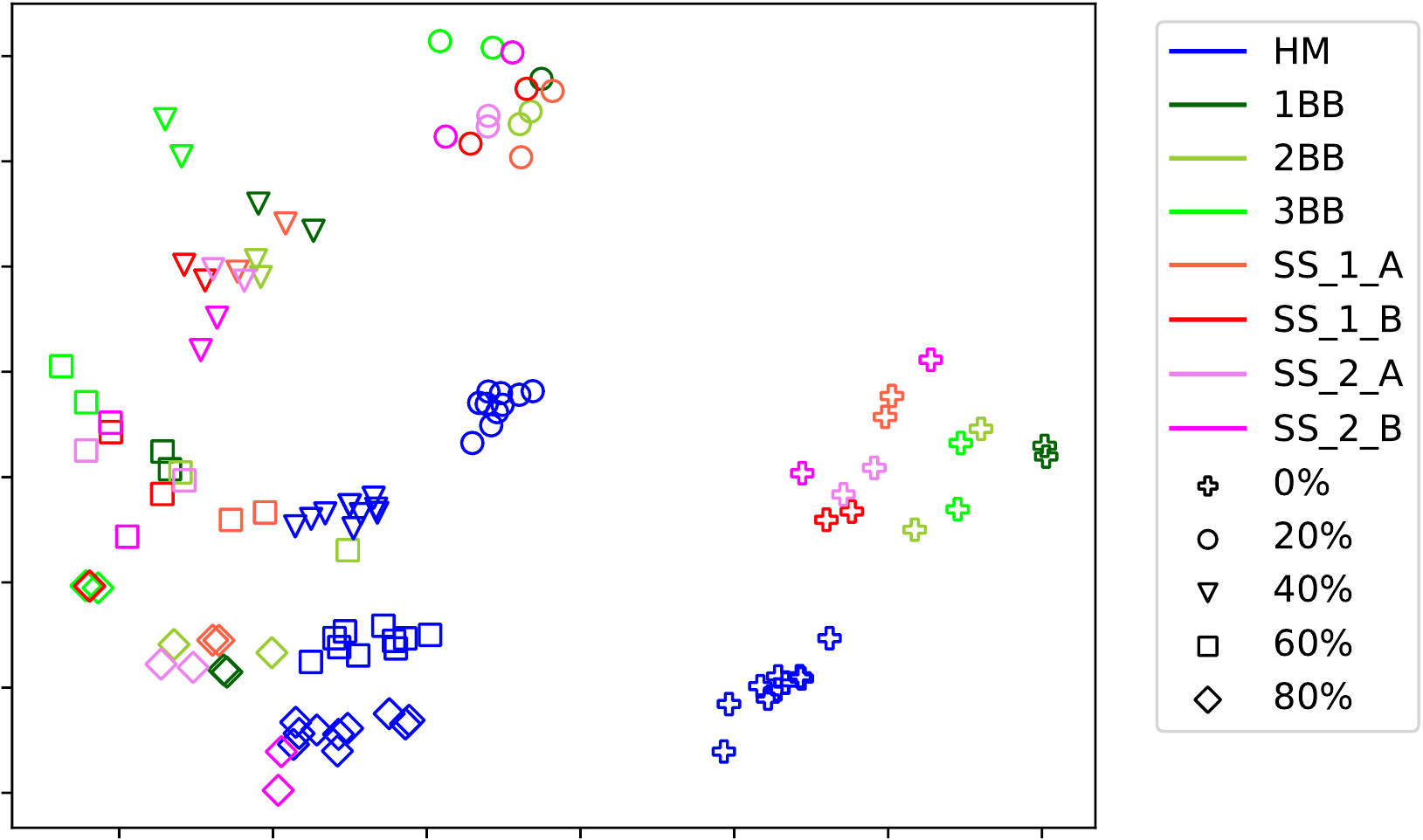}
         \caption{FPCA over the current signals.}
     \end{subfigure}
     \hfill
     \begin{subfigure}[htp]{0.48\textwidth}
         \centering
         \includegraphics[width=\textwidth,height=1.5in]{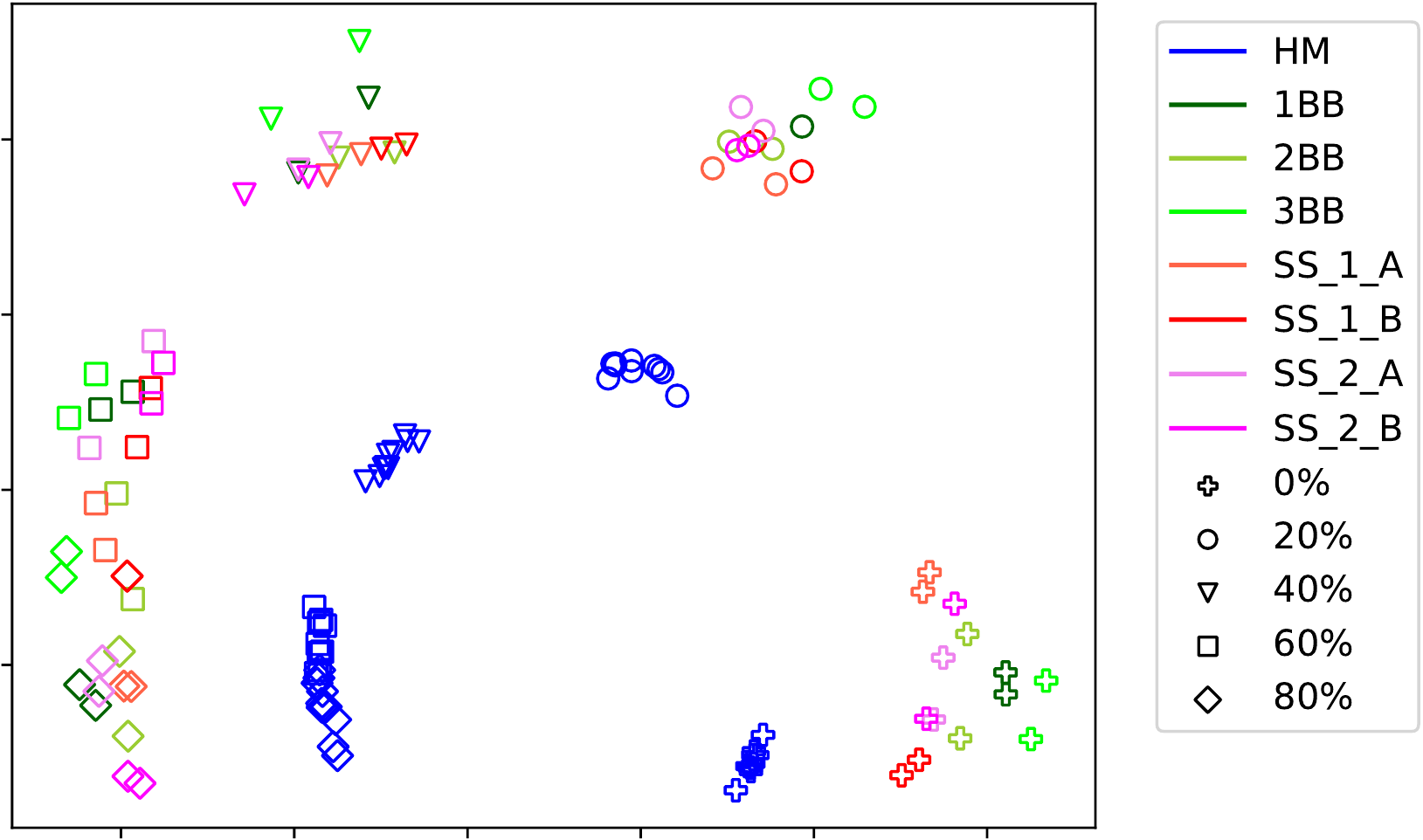}
         \caption{FPCA over the current signal derivatives.}
     \end{subfigure}
     \\
     \begin{subfigure}[htp]{0.48\textwidth}
         \centering
         \includegraphics[width=\textwidth,height=1.5in]{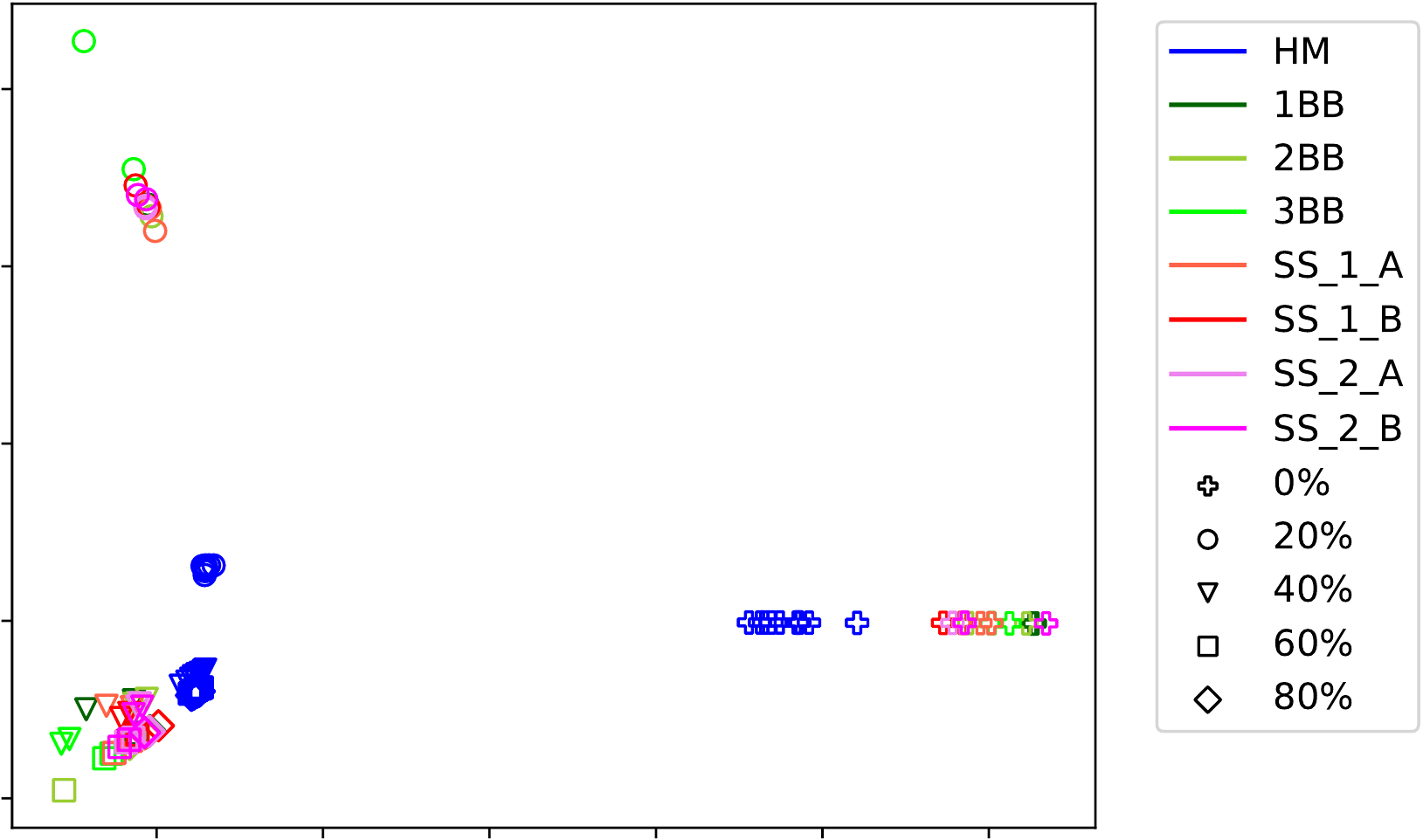}
         \caption{FDM over the current signals.}
     \end{subfigure}
     \hfill
     \begin{subfigure}[htp]{0.48\textwidth}
         \centering
         \includegraphics[width=\textwidth,height=1.5in]{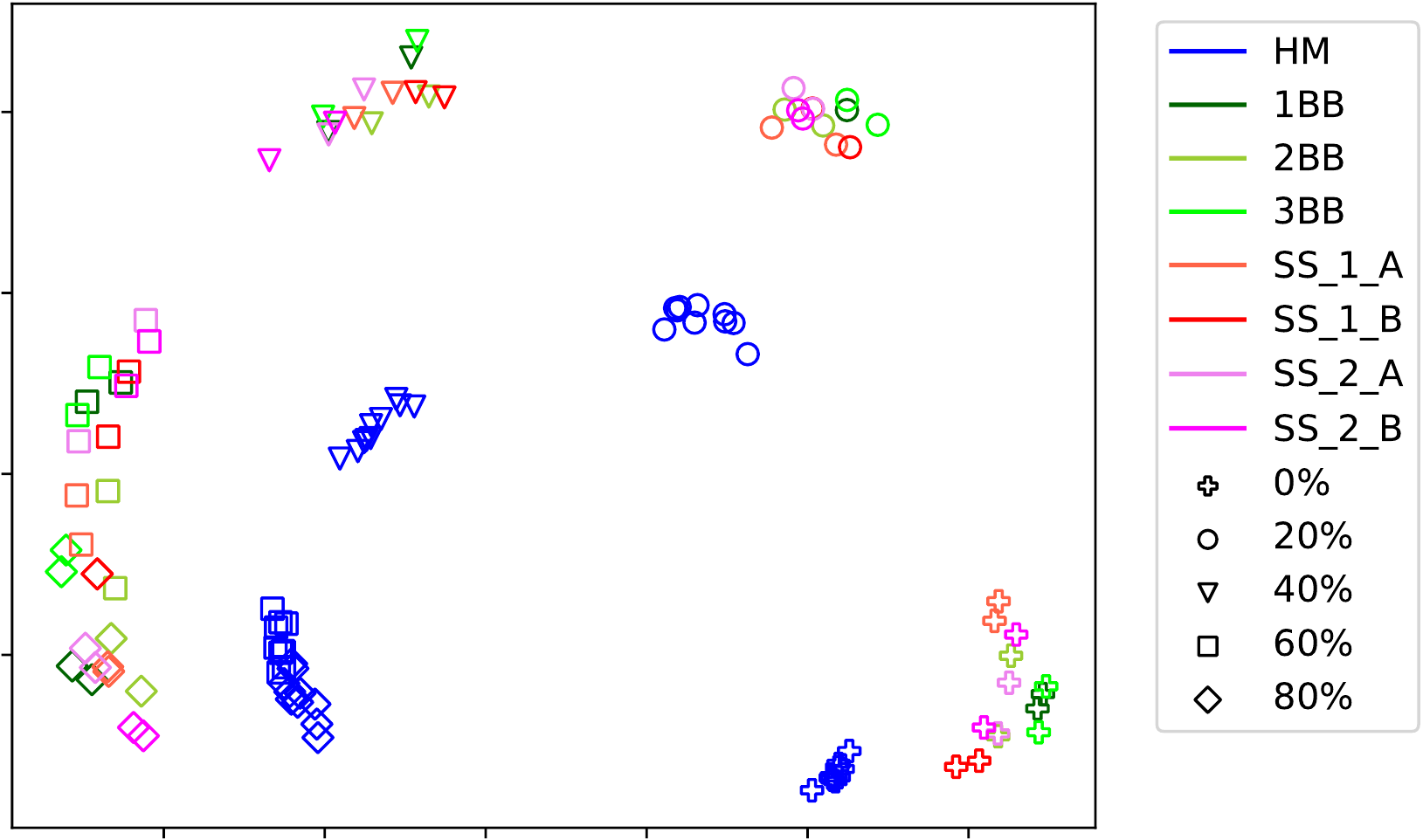}
         \caption{FDM over the current signal derivatives.}
     \end{subfigure}
        \caption{FPCA and FDM embedding over the current signals and their derivatives.}
    \label{FIG:current-embedding-signals}
\end{figure}

The goal of the first experiment is to detect motors with faults using dimension reduction methods over the current signals and their derivatives.
Figure~\ref{FIG:current-embedding-signals} shows the scatterplots of FPCA and FDM scores over them. 

While the FPCA embedding over the current signals fails for data with high load percentages, the FPCA embedding obtained for the current signal derivatives groups data from the healthy motor and data from the motors with faults into separable clusters.
Therefore, there is more discriminatory information in the growth of the current data than in the amplitude.
The influence of load on the embeddings can also be observed, as the scores are grouped by load instead of by type of failure.
Even so, it is worth highlighting the scores corresponding to motors with \num{3} broken bars, which appear at the embedding edges.  

Similar analysis can be applied to the FDM embeddings, particularly to the one obtained using current derivatives, which groups data from the healthy motor in a distinct concentric circle, well separated from the rest.
On the other hand, the embedding over the current signals shows an improvement compared to the FPCA embedding for this type of data, as it avoids mixing data from faulty motors with data from the healthy one.

In summary, we have obtained an embedding that centralizes all the components from the healthy motor and excludes those from faulty motors.
This is a novel contribution to the state of the art in induction motor faults detection using current instead of its signature and obtained in an automated way.
This suggests that, by using an unsupervised analysis, it is possible to identify data coming from faulty motors.
However, accurately identifying the type of malfunction remains a challenging task that requires additional information to be achieved.
In the subsequent experiments, we will only use data from faulty motors in order to distinguish the type of fault.

\subsubsection{Analysis of the current signature}

\begin{figure}[t]
    \begin{subfigure}{0.48\textwidth}
         \centering
         \includegraphics[width=\textwidth,height=1.5in]{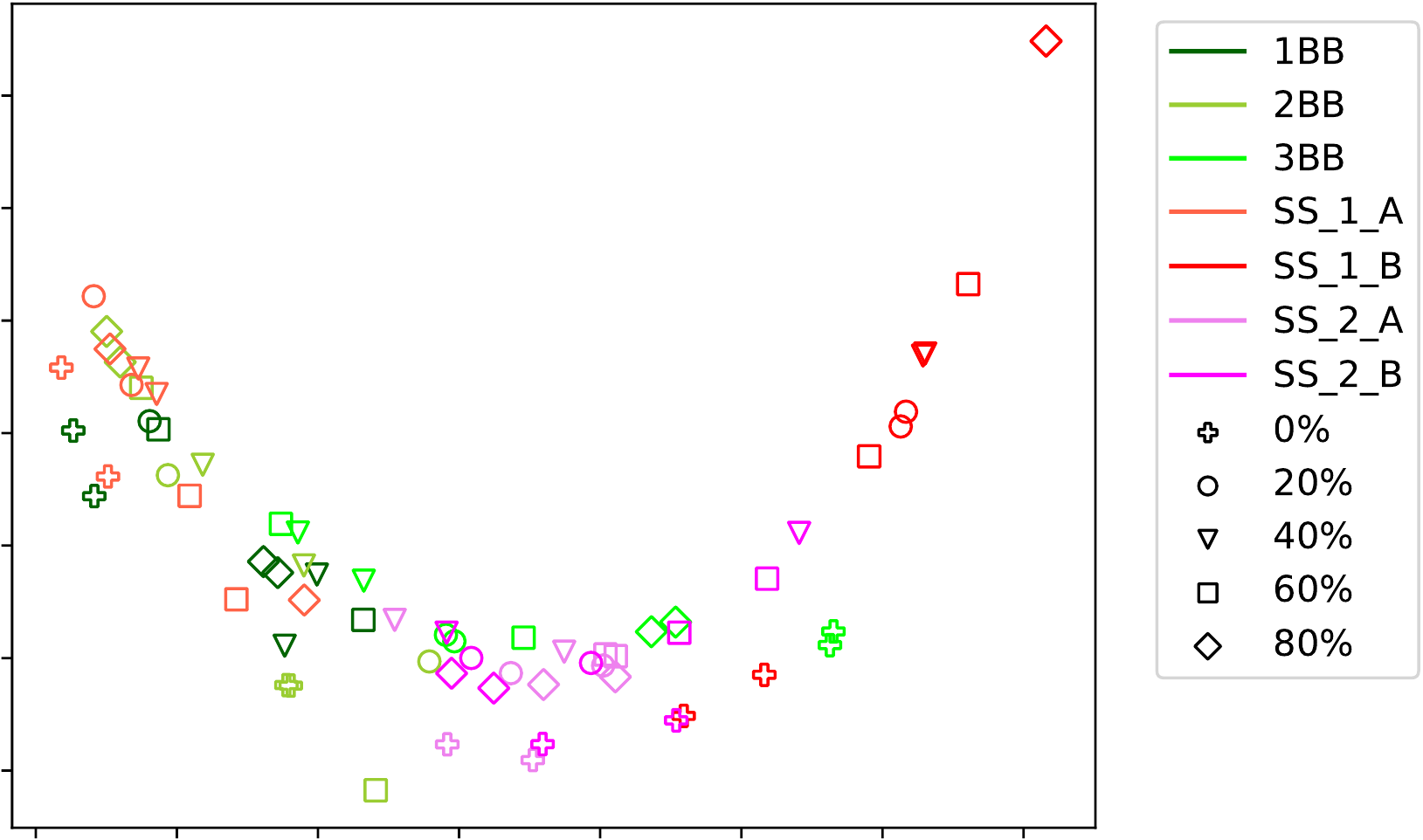}
         \caption{FPCA over the current signatures.}
     \end{subfigure}
     \hfill
     \begin{subfigure}{0.48\textwidth}
         \centering
         \includegraphics[width=\textwidth,height=1.5in]{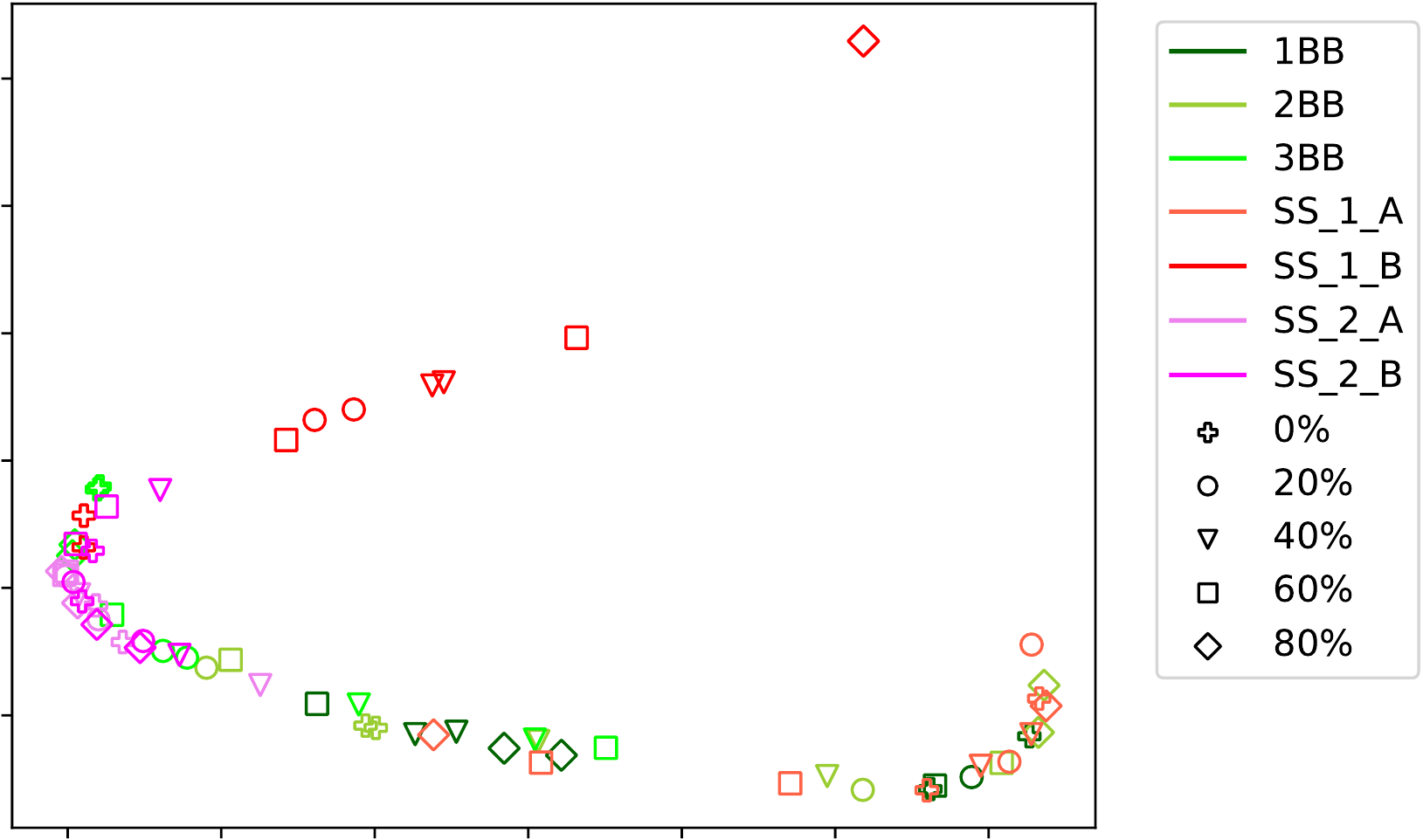}
         \caption{FDM over the current signatures.}
     \end{subfigure}
    \caption{FPCA and FDM embedding over the current signatures.}
    \label{FIG:current-embedding-signatures}
\end{figure}

In the second experiment, we will verify if the proposed dimensionality reduction techniques allow us to determine the type of fault present in the motors.
We will examine whether these techniques can replace the common visual analysis of the current signature in induction motor fault detection literature~\cite{Faiz}.
Figure~\ref{FIG:current-embedding-signatures} shows the scatterplots of FPCA and FDM scores over the current signatures.

Both FPCA and FDM embeddings separate data from motors with low-frequency load oscillations into three groups: motors with load oscillations at \SI{2}{\hertz}; motors with loads at \SI{1}{\hertz}/\SI{1}{\milli\volt}; and motors with loads oscillations at \SI{1}{\hertz}/\SI{1.5}{\milli\volt}. 
However, signals from motors with broken bars appear scattered. 
Therefore, any embedding makes it possible to diagnose faults due to broken bars from faults due to low-frequency load oscillations.

In conclusion, dimensionality reduction methods applied to the current signature are not sufficient to diagnose the type of failure in induction motors.
We need to use more information and we resort to the instantaneous active power.
In the same way as in the previous experiments, we will analyze the power signal, its derivative and finally, its signature.

\subsubsection{Analysis of the instantaneous active power signal and its derivative}

\begin{figure}[t]
\centering
     \begin{subfigure}{0.48\textwidth}
         \centering
         \includegraphics[width=\textwidth,height=1.5in]{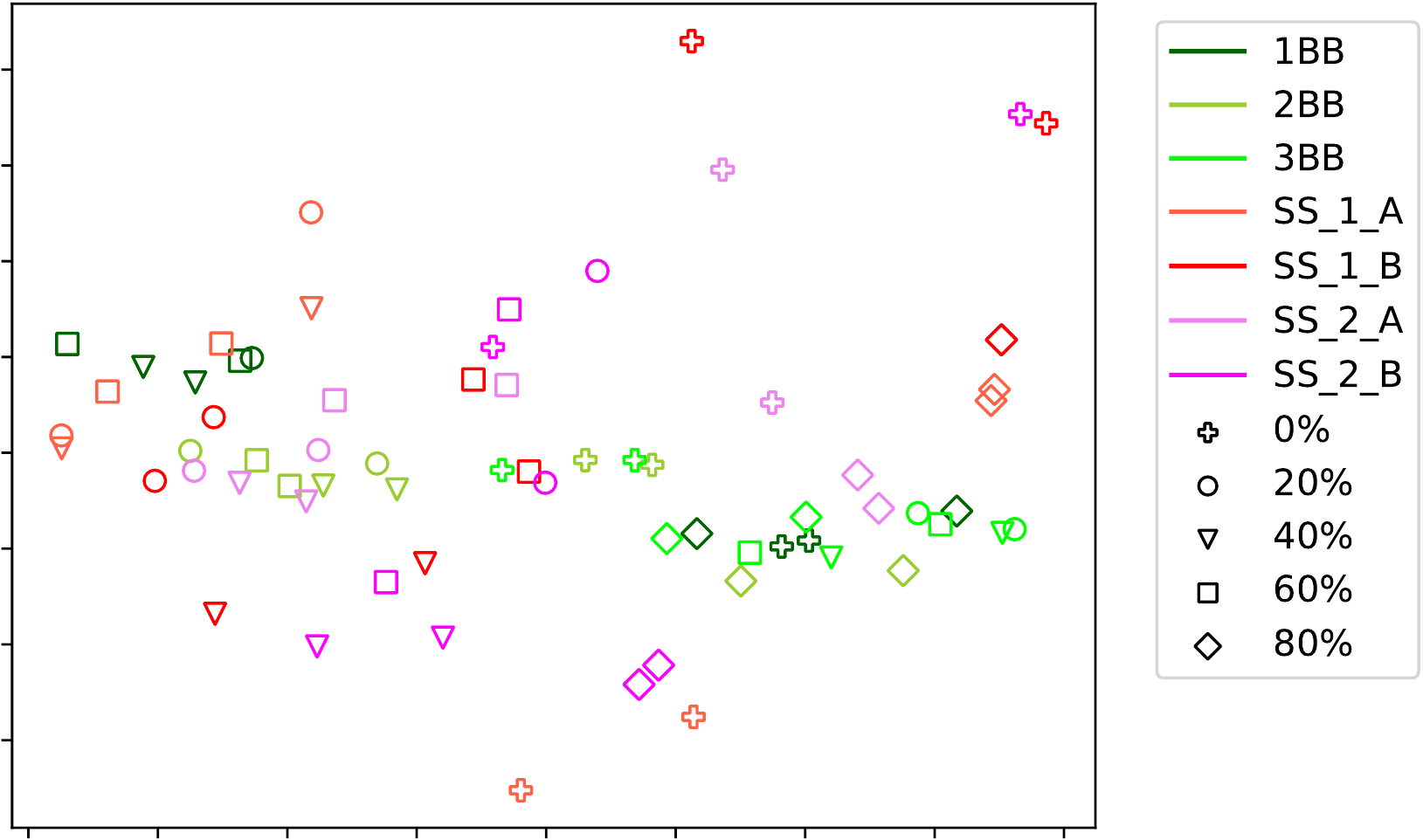}
         \caption{FPCA over the IAP signals.}
     \end{subfigure}
     \hfill
     \begin{subfigure}{0.48\textwidth}
         \centering
         \includegraphics[width=\textwidth,height=1.5in]{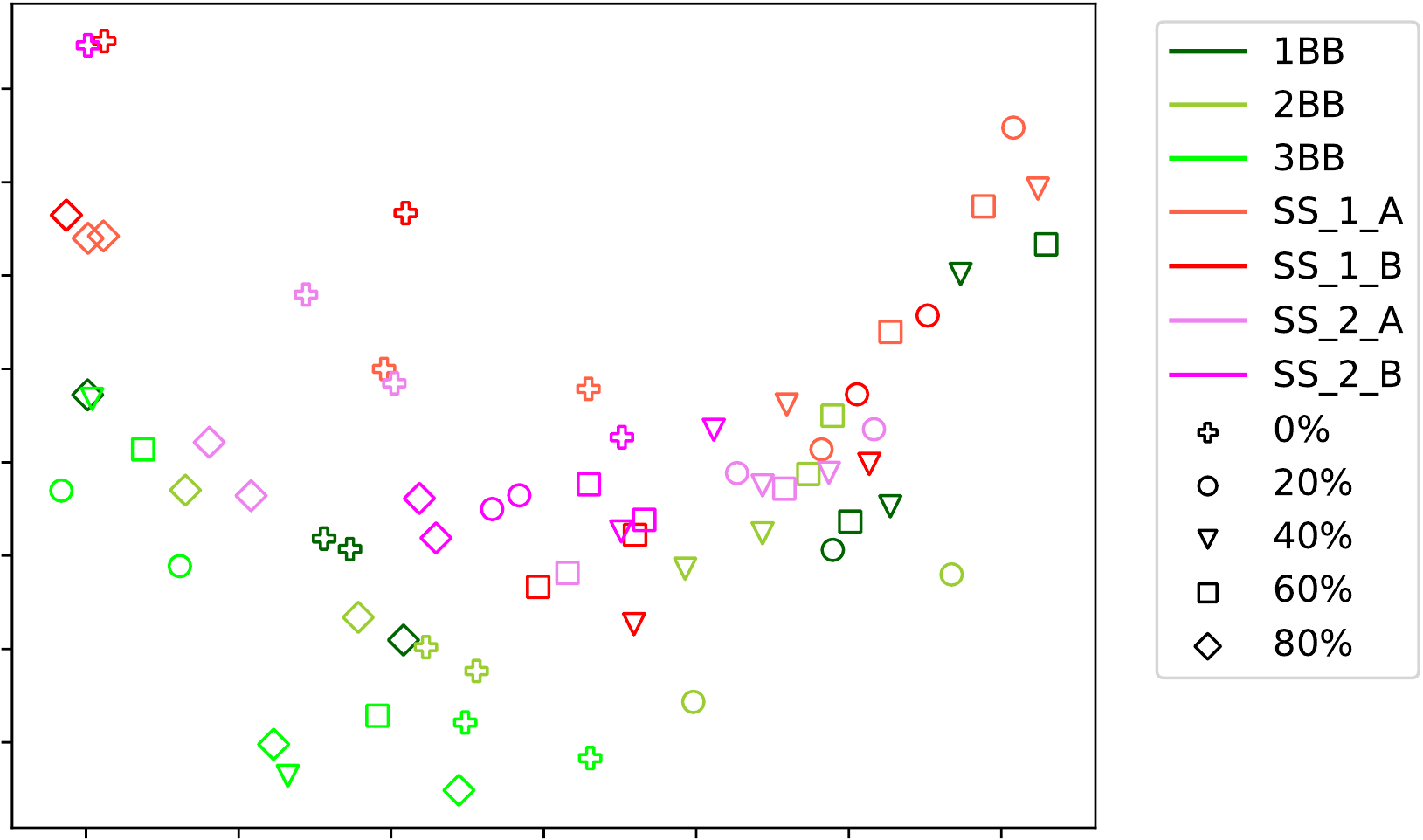}
         \caption{FPCA over the IAP signal derivatives.}
     \end{subfigure}
     \\
     \begin{subfigure}{0.48\textwidth}
         \centering
         \includegraphics[width=\textwidth,height=1.5in]{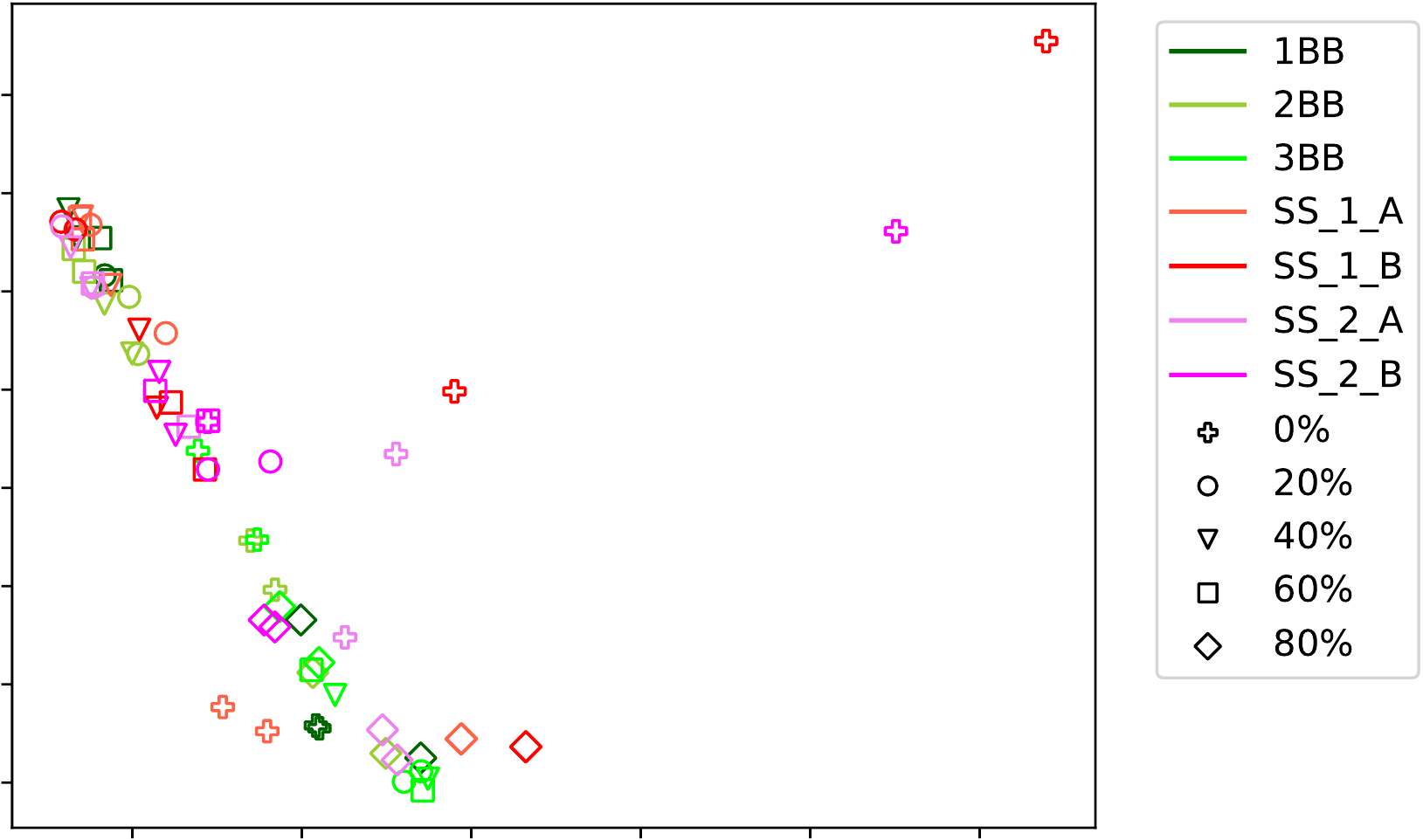}
         \caption{FDM over the IAP signals.}
     \end{subfigure}
     \hfill
     \begin{subfigure}{0.48\textwidth}
         \centering
         \includegraphics[width=\textwidth,height=1.5in]{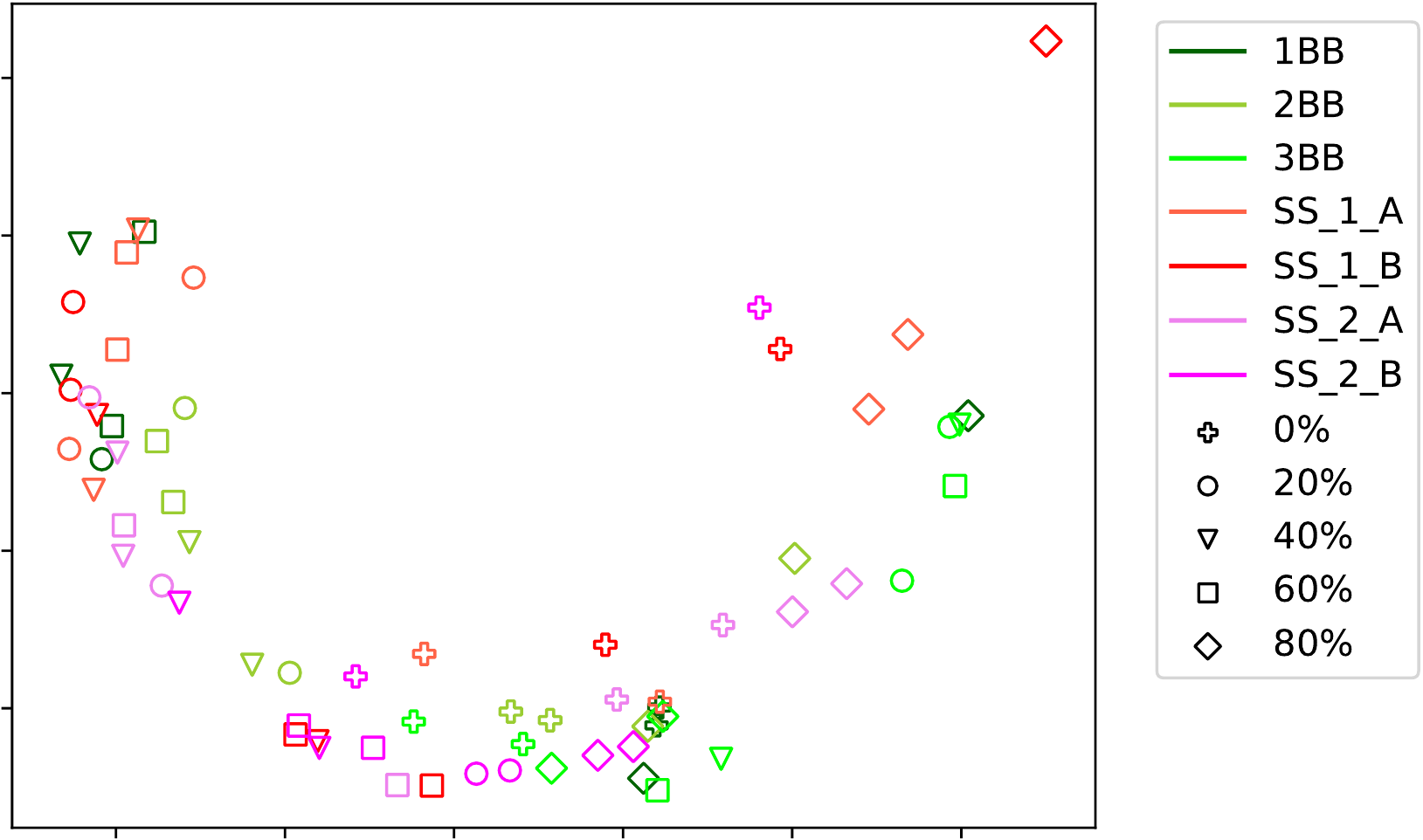}
         \caption{FDM over the IAP signal derivatives.}
     \end{subfigure}
        \caption{FPCA and FDM embedding over the IAP signals and their derivatives.}
    \label{FIG:power-embedding-signals}
\end{figure}

In the third experiment, the instantaneous active power signals and their derivatives are analyzed.
Figure~\ref{FIG:power-embedding-signals} shows the scatterplots of FPCA and FDM scores over the IAP signals and their derivatives.

None of the obtained embeddings allows to diagnose the type of fault. 
However, we can stand out the one obtained by FPCA on the IAP signal derivatives as it is the one that presents data more separated, grouped from motors with faults due to broken bars at the bottom, data from faulty motors with load oscillations at \SI{2}{\hertz} in the middle, and data from faulty motors with load oscillations at \SI{1}{\hertz} at the extremes.
Nevertheless, the clusters appear very close and even overlapped.

\subsubsection{Analysis of the instantaneous active power signature}

\begin{figure}[htp]
    \begin{subfigure}{0.48\textwidth}
         \centering
         \includegraphics[width=\textwidth,height=1.5in]{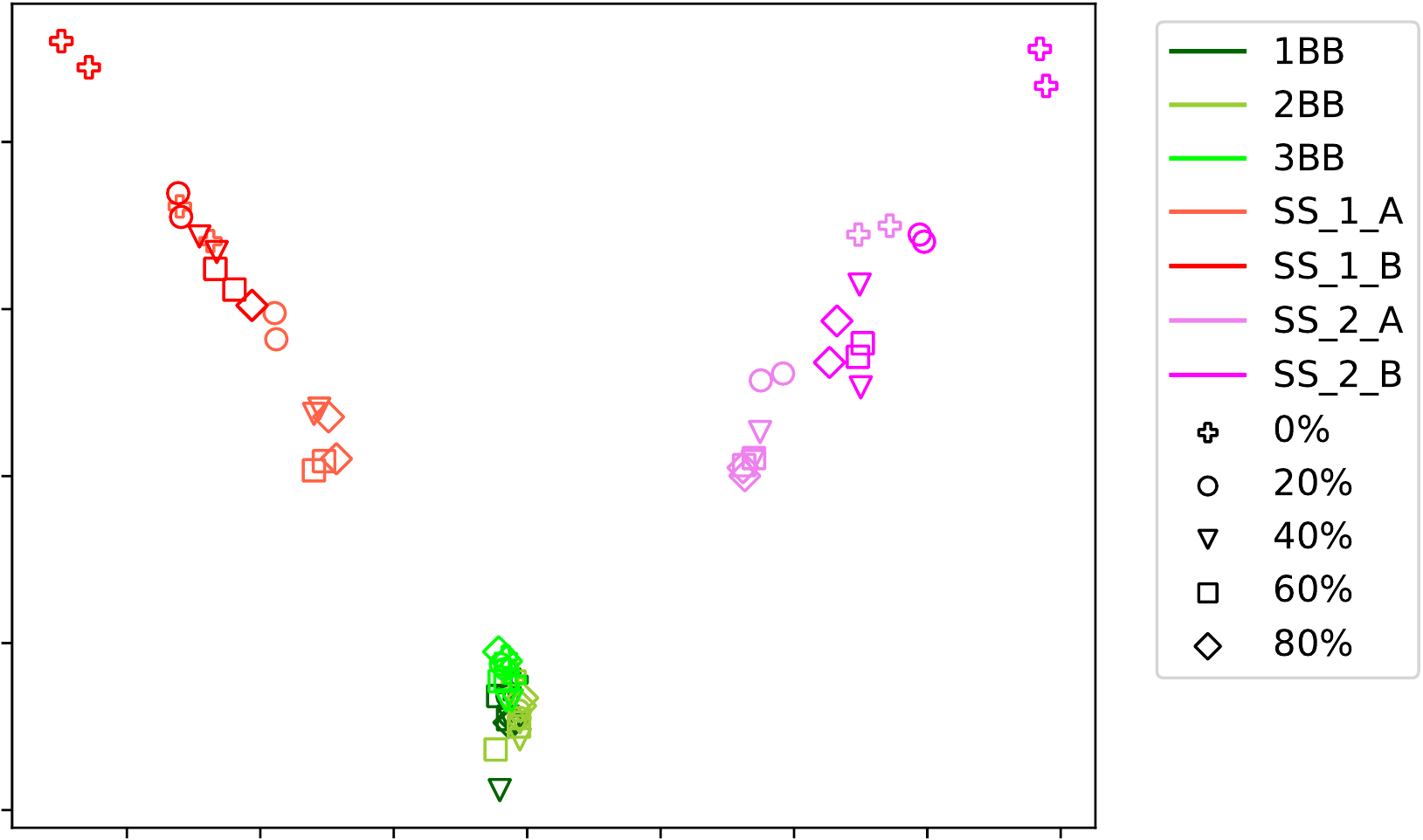}
         \caption{FPCA over the IAP signatures.}
     \end{subfigure}
     \hfill
     \begin{subfigure}{0.48\textwidth}
         \centering
         \includegraphics[width=\textwidth,height=1.5in]{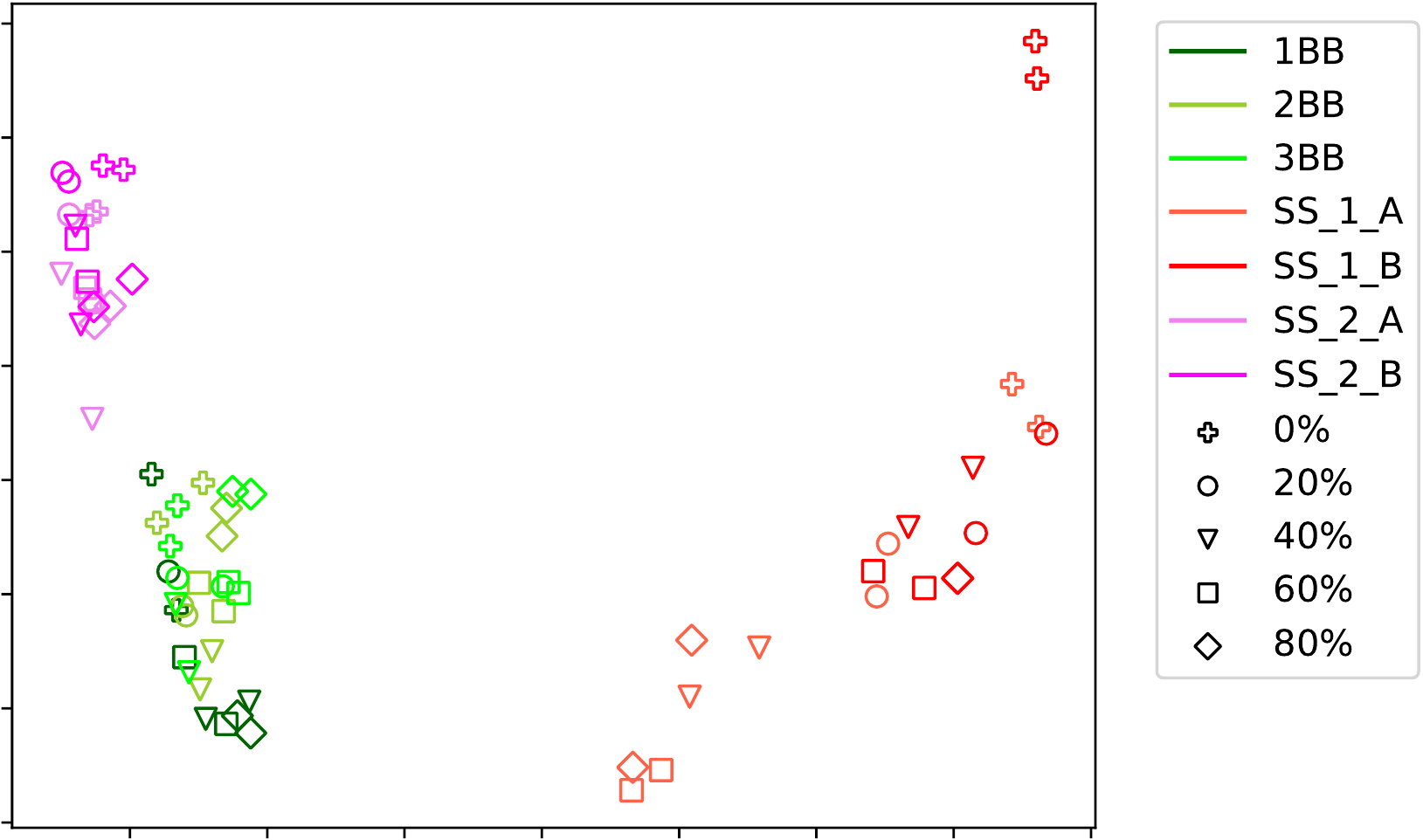}
         \caption{FDM over the IAP signatures.}
     \end{subfigure}
    \caption{FPCA and FDM embedding over the IAP signatures.}
    \label{FIG:power-embedding-signatures}
\end{figure}

After the last experiment, we will try to diagnose the type of motor fault by analyzing the IAP signatures.
The FPCA and FDM embeddings obtained are displayed in Figure~\ref{FIG:power-embedding-signatures}. 

FPCA allows to differentiate motors with failures due to broken bars from motors with failures due to load oscillations, distinguishing those at 
\SI{1}{\hertz} and \SI{1.5}{\hertz}, since they appear well separated in the embedding.
However, detecting the number of broken bars of the motors remains a challenge as they appear grouped together in the embedding.
Another important contribution is that by using the active power signature, the embedding obtained gives much less importance to the motor load percentage than in the rest of the embeddings.
By comparison, FDM embedding also keeps the above classes separated, but the separation distance between classes is smaller.

\subsection{Detection and diagnosis algorithm}
\label{subsec:scheme}

\begin{figure}[htp]
    \centering
    \includegraphics[scale=0.7]{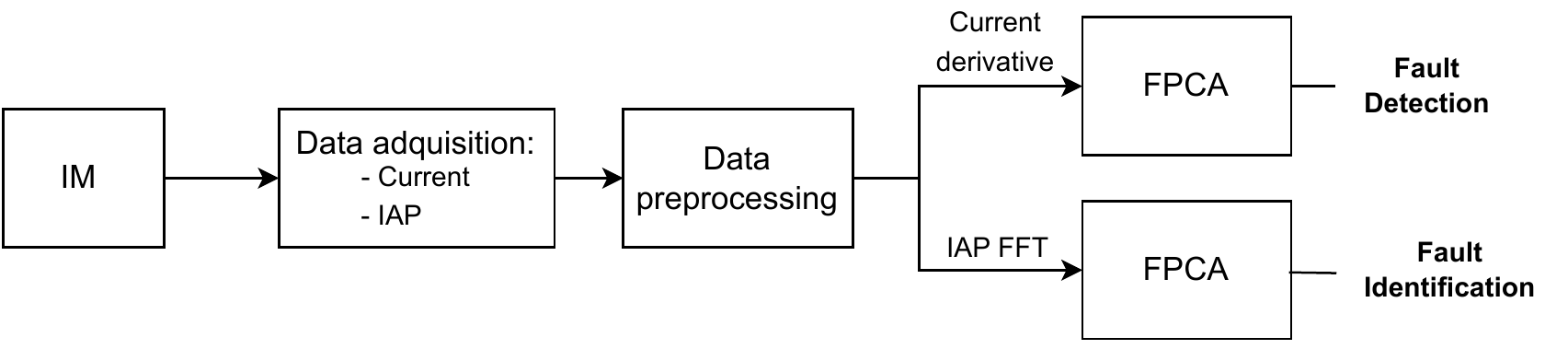}
    \caption{Unsupervised detection and diagnosis algorithm.}
    \label{FIG:algorithm}
\end{figure}

Based on the previous results, it can be concluded that functional dimensionality reduction methods offer a good performance for electrical signals from induction motors, which enables the proposal of an algorithm for detecting and diagnosing induction faults.
Figure~\ref{FIG:algorithm} depicts the proposed induction faults detection algorithm.

First of all, current signals are collected from an induction motor, along with line voltage signals, which facilitate the computation of instantaneous active power data.
Then, a preprocessing step consisting of aligning data, scaling to the range of \num{-1} and \num{1}, and keeping only the first \num{750} steps as a representative sample is performed.
Consecutively, FPCA is applied to the time-domain derivative of a single stator current in order to distinguish healthy motors from motors with faults.
Next, in the event that the motor exhibits faults, FPCA is applied to the frequency domain, specifically to the frequency spectrum of the instantaneous active power signal, to distinguish between motors with broken bars faults and those with low-frequency load oscillation faults, as well as their various subtypes.

\section{Conclusions}
\label{sec:Concl}

Implementing strategies to detect and diagnose early faults in rotating electrical machines online is crucial for ensuring the reliability and safety of modern industrial systems.
Its execution allows planning interruptions of continuous production processes in scheduled stops, thus reducing maintenance time and associated economic losses~\cite{mobley2004maintenance, etde_21044166}.
The diagnosis of faults present in an REM is integrated by the detection, identification and isolation of an anomaly, which can be applied by using the information obtained on the state of operation of the equipment or drive~\cite{Tavner1621470}.
As a result, it is possible to consider fault diagnosis as a pattern recognition problem with respect to the condition of an REM~\cite{1546063}. 

The proposed unsupervised scheme consist of using functional dimensionality reduction techniques, specifically FPCA and FDM, to detect and identify the presence of faults due to broken bars and low-frequency load oscillations in induction motors.
An analysis of a single stator current and an analysis of the instantaneous active power of the IM is carried out.
In this analysis, both the raw data and its derivatives, as well as their signatures, are used.

The results obtained from the scheme proposed are very encouraging, revealing a potential use in the future not only for real-time detection of the presence of a fault in an IM, but also in the identification of a greater number of types of faults present through an offline analysis.
Both FPCA and FDM give very similar results, although FPCA is more competitive for these data. We have seen that FPCA on the derivative of the current signal allows to distinguish motors with faults from healthy motors and applying this technique for the instantaneous active power signature it is possible to diagnose the type of motor failure: broken bars or low-frequency load oscillations.

\section*{Acknowledgements}

The authors acknowledge financial support from the European Regional Development Fund and the Spanish State Research Agency of the Ministry of Economy, Industry, and Competitiveness under the project PID2019-106827GB-I00 / AEI / 10.13039/501100011033.
They also thank the UAM--ADIC Chair for Data Science and Machine Learning.

\bibliography{refs}

\begin{thebibliography}{10}
\expandafter\ifx\csname url\endcsname\relax
  \def\url#1{\texttt{#1}}\fi
\expandafter\ifx\csname urlprefix\endcsname\relax\def\urlprefix{URL }\fi
\expandafter\ifx\csname href\endcsname\relax
  \def\href#1#2{#2} \def\path#1{#1}\fi

\bibitem{mobley2004maintenance}
R.~Mobley, \href{https://books.google.es/books?id=-adqlAEACAAJ}{Maintenance
  Fundamentals}, Plant Engineering Series, Elsevier/Butterworth Heinemann,
  2004.
\newline\urlprefix\url{https://books.google.es/books?id=-adqlAEACAAJ}

\bibitem{etde_21044166}
R.~Isermann, Fault-diagnosis systems. an introduction from fault detection to
  fault tolerance (Jul 2006).
\newblock \href {https://doi.org/10.1007/3-540-30368-5}
  {\path{doi:10.1007/3-540-30368-5}}.

\bibitem{Tavner1621470}
P.~Tavner, L.~Ran, J.~Penman, H.~Sedding,
  \href{http://cds.cern.ch/record/1621470}{{Condition monitoring of rotating
  electrical machines}}, The Institution of Engineering and Technology,
  Stevenage, 2008.
\newline\urlprefix\url{http://cds.cern.ch/record/1621470}

\bibitem{1546063}
S.~Nandi, H.~Toliyat, X.~Li, Condition monitoring and fault diagnosis of
  electrical motors—a review, IEEE Transactions on Energy Conversion 20~(4)
  (2005) 719--729.
\newblock \href {https://doi.org/10.1109/TEC.2005.847955}
  {\path{doi:10.1109/TEC.2005.847955}}.

\bibitem{GANGSAR2020106908}
P.~Gangsar, R.~Tiwari,
  \href{https://www.sciencedirect.com/science/article/pii/S0888327020302946}{Signal
  based condition monitoring techniques for fault detection and diagnosis of
  induction motors: A state-of-the-art review}, Mechanical Systems and Signal
  Processing 144 (2020) 106908.
\newblock \href {https://doi.org/https://doi.org/10.1016/j.ymssp.2020.106908}
  {\path{doi:https://doi.org/10.1016/j.ymssp.2020.106908}}.
\newline\urlprefix\url{https://www.sciencedirect.com/science/article/pii/S0888327020302946}

\bibitem{LIU201833}
R.~Liu, B.~Yang, E.~Zio, X.~Chen,
  \href{https://www.sciencedirect.com/science/article/pii/S0888327018300748}{Artificial
  intelligence for fault diagnosis of rotating machinery: A review}, Mechanical
  Systems and Signal Processing 108 (2018) 33--47.
\newblock \href {https://doi.org/https://doi.org/10.1016/j.ymssp.2018.02.016}
  {\path{doi:https://doi.org/10.1016/j.ymssp.2018.02.016}}.
\newline\urlprefix\url{https://www.sciencedirect.com/science/article/pii/S0888327018300748}

\bibitem{wang2016functional}
J.-L. Wang, J.-M. Chiou, H.-G. M{\"u}ller, Functional data analysis, Annual
  Review of Statistics and its application 3 (2016) 257--295.

\bibitem{Shang2014ASO}
H.~L. Shang, A survey of functional principal component analysis, AStA Advances
  in Statistical Analysis 98 (2014) 121--142.

\bibitem{barroso2023functional}
M.~Barroso, C.~M. Alaíz, Ángela Fernández, J.~L. Torrecilla, Functional
  diffusion maps (2023).
\newblock \href {http://arxiv.org/abs/2304.14378} {\path{arXiv:2304.14378}}.

\bibitem{ReviewAlgorithmInductionMotors}
P.~Kumar, A.~Hati, Review on machine learning algorithm based fault detection
  in induction motors, Archives of Computational Methods in Engineering 28 (06
  2020).
\newblock \href {https://doi.org/10.1007/s11831-020-09446-w}
  {\path{doi:10.1007/s11831-020-09446-w}}.

\bibitem{app11062761}
K.~Kudelina, T.~Vaimann, B.~Asad, A.~Rassõlkin, A.~Kallaste, G.~Demidova,
  \href{https://www.mdpi.com/2076-3417/11/6/2761}{Trends and challenges in
  intelligent condition monitoring of electrical machines using machine
  learning}, Applied Sciences 11~(6) (2021).
\newblock \href {https://doi.org/10.3390/app11062761}
  {\path{doi:10.3390/app11062761}}.
\newline\urlprefix\url{https://www.mdpi.com/2076-3417/11/6/2761}

\bibitem{MethodsConditionDetectionMachines}
K.~Kudelina, B.~Asad, T.~Vaimann, A.~Rassõlkin, A.~Kallaste, H.~V. Khang,
  Methods of condition monitoring and fault detection for electrical machines,
  Energies 14 (2021) 7459.
\newblock \href {https://doi.org/10.3390/en14227459}
  {\path{doi:10.3390/en14227459}}.

\bibitem{BRITO2022108105}
L.~C. Brito, G.~A. Susto, J.~N. Brito, M.~A. Duarte,
  \href{https://www.sciencedirect.com/science/article/pii/S0888327021004891}{An
  explainable artificial intelligence approach for unsupervised fault detection
  and diagnosis in rotating machinery}, Mechanical Systems and Signal
  Processing 163 (2022) 108105.
\newblock \href {https://doi.org/https://doi.org/10.1016/j.ymssp.2021.108105}
  {\path{doi:https://doi.org/10.1016/j.ymssp.2021.108105}}.
\newline\urlprefix\url{https://www.sciencedirect.com/science/article/pii/S0888327021004891}

\bibitem{AdvancedSignalConditionMonitoring}
R.~Jaros, R.~Byrtus, J.~Dohnal, L.~Danys, J.~Baroš, J.~Koziorek, P.~Zmij,
  R.~Martinek, Advanced signal processing methods for condition monitoring,
  Archives of Computational Methods in Engineering (10 2022).
\newblock \href {https://doi.org/10.1007/s11831-022-09834-4}
  {\path{doi:10.1007/s11831-022-09834-4}}.

\bibitem{DeterminationBrokenPCA}
P.~Moallem, B.~Mirzaeian~Dehkordi, M.~Shirvani, Determination of the number of
  broken rotor bars in squirrel-cage induction motors using wavelet, pca and
  neural networks, International Review of Electrical Engineering 4 (2009)
  242--248.

\bibitem{7544203}
V.~F. Pires, J.~F. Martins, A.~J. Pires, L.~Rodrigues, Induction motor broken
  bar fault detection based on mcsa, mscsa and pca: A comparative study, in:
  2016 10th International Conference on Compatibility, Power Electronics and
  Power Engineering (CPE-POWERENG), 2016, pp. 298--303.
\newblock \href {https://doi.org/10.1109/CPE.2016.7544203}
  {\path{doi:10.1109/CPE.2016.7544203}}.

\bibitem{7049012}
M.~O. Mustafa, G.~Georgoulas, G.~Nikolakopoulos, Principal component analysis
  anomaly detector for rotor broken bars, in: IECON 2014 - 40th Annual
  Conference of the IEEE Industrial Electronics Society, 2014, pp. 3462--3467.
\newblock \href {https://doi.org/10.1109/IECON.2014.7049012}
  {\path{doi:10.1109/IECON.2014.7049012}}.

\bibitem{manifold_learning}
L.~Cayton, Algorithms for manifold learning, Univ. of California at San Diego
  Tech. Rep 12~(1-17) (2005) 1.

\bibitem{Unsupervised_Functional_Data_Analysis}
M.~Herrmann, F.~Scheipl, \href{https://arxiv.org/abs/2012.11987}{Unsupervised
  functional data analysis via nonlinear dimension reduction} (2020).
\newblock \href {https://doi.org/10.48550/ARXIV.2012.11987}
  {\path{doi:10.48550/ARXIV.2012.11987}}.
\newline\urlprefix\url{https://arxiv.org/abs/2012.11987}

\bibitem{Benchmarking_time_series_classification}
F.~Pfisterer, L.~Beggel, X.~Sun, F.~Scheipl, B.~Bischl, Benchmarking time
  series classification -- functional data vs machine learning approaches
  (2019).
\newblock \href {https://doi.org/10.48550/ARXIV.1911.07511}
  {\path{doi:10.48550/ARXIV.1911.07511}}.

\bibitem{Nonlinear_manifold_representations_for_functional_data}
D.~Chen, H.-G. Müller, \href{https://doi.org/10.1214%2F11-aos936}{Nonlinear
  manifold representations for functional data}, The Annals of Statistics
  40~(1) (feb 2012).
\newblock \href {https://doi.org/10.1214/11-aos936}
  {\path{doi:10.1214/11-aos936}}.
\newline\urlprefix\url{https://doi.org/10.1214%2F11-aos936}

\bibitem{abdi2010principal}
H.~Abdi, L.~J. Williams, Principal component analysis, Wiley interdisciplinary
  reviews: computational statistics 2~(4) (2010) 433--459.

\bibitem{Ramsay_FDA}
J.~O. Ramsay, B.~W. Silverman,
  \href{http://www.worldcat.org/isbn/9780387400808}{Functional Data Analysis},
  Springer, 2005.
\newline\urlprefix\url{http://www.worldcat.org/isbn/9780387400808}

\bibitem{Diffusion_Maps}
R.~R. Coifman, S.~Lafon,
  \href{https://www.sciencedirect.com/science/article/pii/S1063520306000546}{Diffusion
  maps}, Applied and Computational Harmonic Analysis 21~(1) (2006) 5--30,
  special Issue: Diffusion Maps and Wavelets.
\newblock \href {https://doi.org/https://doi.org/10.1016/j.acha.2006.04.006}
  {\path{doi:https://doi.org/10.1016/j.acha.2006.04.006}}.
\newline\urlprefix\url{https://www.sciencedirect.com/science/article/pii/S1063520306000546}

\bibitem{An_Introduction_to_Diffusion_Maps}
J.~De~la Porte, B.~Herbst, W.~Hereman, S.~Van Der~Walt, An introduction to
  diffusion maps, in: Proceedings of the 19th symposium of the pattern
  recognition association of South Africa (PRASA 2008), Cape Town, South
  Africa, 2008, pp. 15--25.

\bibitem{Bossio5071292}
G.~R. Bossio, C.~H. De~Angelo, J.~M. Bossio, C.~M. Pezzani, G.~O. Garcia,
  Separating broken rotor bars and load oscillations on im fault diagnosis
  through the instantaneous active and reactive currents, IEEE Transactions on
  Industrial Electronics 56~(11) (2009) 4571--4580.
\newblock \href {https://doi.org/10.1109/TIE.2009.2024656}
  {\path{doi:10.1109/TIE.2009.2024656}}.

\bibitem{derivative_finite_difference}
M.~Renardy, R.~Rogers, \href{https://books.google.es/books?id=0RmDjrFO3CUC}{An
  Introduction to Partial Differential Equations}, Texts in Applied
  Mathematics, Springer New York, 2004.
\newline\urlprefix\url{https://books.google.es/books?id=0RmDjrFO3CUC}

\bibitem{fft}
J.~Cooley, J.~Tukey, An algorithm for the machine calculation of complex
  fourier series, Mathematics of Computation 19~(90) (1965) 297--301.

\bibitem{Faiz}
J.~Faiz, V.~Ghorbanian, G.~Joksimovic, Fault Diagnosis of Induction Motors, The
  Institution of Engineering and Technology, 2017.
\newblock \href {https://doi.org/10.1049/PBPO108E}
  {\path{doi:10.1049/PBPO108E}}.

\end{thebibliography}

\end{document}